\documentclass[
 twocolumn,
 superscriptaddress,
 showpacs,
 amsmath,
 amssymb,
 aps,
 pra,
]{revtex4-1}

\usepackage[dvipdfmx]{graphicx}
\usepackage{dcolumn}
\usepackage{bm}

\usepackage{algcompatible}
\usepackage[size=small]{caption}
\DeclareCaptionFormat{algorithm}{\hrulefill\vskip-8pt\hrulefill\par#1#2#3\vskip-4pt\hrulefill}
\usepackage{etoolbox}
\AtEndEnvironment{algorithm}{\noindent\hrulefill\par\nobreak\vskip-8pt\hrulefill}
\usepackage{newfloat}
\DeclareFloatingEnvironment[
  fileext=loa,
  listname=List of Algorithms,
  name=ALGORITHM,
  placement=tbhp,
]{algorithm}
\DeclareMathOperator*{\argmin}{arg\,min}

\usepackage{mathtools}
\usepackage{amsthm}
\newtheorem{theorem}{Theorem}

\usepackage{subcaption}
\usepackage{xcolor}
\usepackage{diagbox}
\usepackage{multirow}
\usepackage{xr}
\usepackage{braket}
\usepackage{amsmath}

\makeatletter
\newcommand{\vast}{\bBigg@{3.0}}
\newcommand{\Vast}{\bBigg@{3.5}}

\makeatother
\allowdisplaybreaks

\begin{document}

\preprint{APS/123-QED}

\title{Quantum Expectation-Maximization Algorithm}

\author{Hideyuki Miyahara}
\email{hideyuki\_miyahara@mist.i.u-tokyo.ac.jp, \\ hmiyahara512@gmail.com}

\affiliation{%
Department of Mathematical Informatics,
Graduate School of Information Science and Technology,
The University of Tokyo,
Tokyo 113-8656, Japan
}%


\author{Kazuyuki Aihara}
\affiliation{%
Institute of Industrial Science, The University of Tokyo,
Tokyo 153-8505, Japan
}%
\affiliation{International Research Center for Neurointelligence (WPI-IRCN), UTIAS,
The University of Tokyo,
Tokyo 113-0033, Japan}

\author{Wolfgang Lechner}
\affiliation{%
Institute for Theoretical Physics, University of Innsbruck, A-6020 Innsbruck, Austria
}%

\date{\today}

\begin{abstract}
Clustering algorithms are a cornerstone of machine learning applications. Recently, a quantum algorithm for clustering based on the $k$-means algorithm has been proposed by Kerenidis, Landman, Luongo and Prakash.
Based on their work, we propose a quantum expectation-maximization (EM) algorithm for Gaussian mixture models (GMMs). The robustness and quantum speedup of the algorithm is demonstrated.
We also show numerically the advantage of GMM over k-means for non-trivial cluster data.
\end{abstract}


\maketitle


\section{Introduction}

Quantum computing has attracted much attention since the discovery of Shor's algorithm~\cite{Shor01, Nielsen02}.
Recently, with the rapid developments in machine learning, physicists have started to consider utilizing quantum computers for machine learning applications \cite{biamonte2017quantum,Lloyd01,schuld2015introduction,Robentrost01,Wiebe01,dunjko2016quantum}.
As a result, quantum machine learning has emerged as an interdisciplinary field between quantum computing and machine learning.
Furthermore, a quantum algorithm for the $k$-means algorithm~\cite{Bishop01, Murphy01} with proven quantum speedup was proposed~\cite{Kerenidis02}.

The $k$-means algorithm is an essential tool in many machine learning applications ~\cite{Bishop01, Murphy01}.
However, the $k$-means algorithm is as a special case of the more general Gaussian mixture model (GMM). In the $k$-means algorithm, each Gaussian has the same weight and the covariance matrix of each Gaussian function is the identity. As a result, the $k$-means algorithm may provide poor estimates of the clusters since the assumptions of the $k$-means algorithm are sometimes too strong to capture all properties of complex data sets.
The expectation-maximization (EM) algorithm~\cite{Dempster01, Bishop01, Murphy01} and variational Bayes (VB) inference~\cite{Bishop01, Murphy01} with the GMM are often used to improve the clustering, since the general GMM can deal with a wider class of data sets. Recently, one of the authors proposed quantum-inspired algorithms for the EM algorithm~\cite{Miyahara03, Miyahara04, Miyahara05} and VB~\cite{Miyahara06}.
In Refs.~\cite{Miyahara05, Miyahara06}, we have succeeded in improving the performances of the EM algorithm and VB. However, the aim of Refs.~\cite{Miyahara05, Miyahara06} is to make use of quantum fluctuations as a numerical tool and not to provide a quantum speedup over a classical algorithm; as a result, the computational costs are almost the same.

In this paper, we propose a quantum algorithm to estimate the parameters of the GMM, which we call the quantum EM (q-EM) algorithm.
To this end, following the spirit of Ref.~\cite{Kerenidis02}, we first introduce a randomized variant of the EM algorithm, i.e. the $\delta$-EM algorithm which includes non-deterministic readout of the data. Then, we formulate a quantum algorithm that realizes a speedup of the $\delta$-EM algorithm with respect to the number of data points.
The q-EM algorithm may be an important step toward quantum machine learning, since the EM algorithm is an essential algorithm in machine learning.

This paper is organized as follows.
In Sec.~\ref{classical-preliminaries-01}, we provide classical preliminaries.
In particular, we review the EM algorithm and then introduce the $\delta$-EM algorithm.
In Sec.~\ref{quantum-EM-01}, we present the detailed procedure of the q-EM algorithm.
Then, in Sec.~\ref{sec-analysis-02-01}, we show its computational cost.
In Sec.~\ref{sec-numerical-simulation}, we show numerical simulations of the $\delta$-EM algorithm to confirm that our starting point is valid.
In Sec.~\ref{disc-01}, we discuss the relationship between the EM algorithm and the $k$-means algorithm.
Finally, Sec.~\ref{conc-01} concludes this paper.

Independently of our work, Iordanis Kerenidis, Alessandro Luongo, and Anupam Prakash proposed an extension of the q-means algorithm to Gaussian mixture models similar to this work using soft clustering~\cite{Kerenidis05}.

\section{Classical preliminaries} \label{classical-preliminaries-01}

In this section, we first review the EM algorithm~\cite{Bishop01, Murphy01} in detail.
We then introduce a randomized variant of the EM algorithm, which we call the $\delta$-EM algorithm.
The purpose of introducing the $\delta$-EM algorithm is the need of a robust variant of the EM algorithm against noise which resembles the non-deterministic quantum measurement of a superposed state.

\subsection{EM algorithm} \label{sec-EM-01}

The EM algorithm is a generic approach to estimate parameters of probability distributions based on maximum likelihood estimation.
For simplicity, we focus on the GMM and review the EM algorithm for the GMM.
Let us consider a $d$-dimensional feature space and assume that we have $N$ data points $\{ y_i \}_{i=1}^N$.
The GMM for $x \in \mathbb{R}^d$ is given by
\begin{align}
  p (y; \theta) &= \sum_{k=1}^K \pi^k \mathcal{N} (y; \mu^k, \Sigma^k),
\end{align}
where $\mathcal{N} (y; \mu^k, \Sigma^k) \coloneqq \frac{1}{(2 \pi)^\frac{d}{2} |\Sigma^k|^\frac{1}{2}} e^{- \frac{1}{2} (y - \mu^k) (\Sigma^k)^{-1} (y - \mu^k)}$ is the $d$-dimensional Gaussian function with mean $\mu^k$ and covariance $\Sigma^k$, and $\sum_k \pi^k = 1$.
To simplify the notation, we define $\theta \coloneqq \{ \pi^k, \mu^k, \Sigma^k \}_{k=1}^K$.

The EM algorithm, which estimates $\theta$, consists of the following two steps which are iterated until convergence.
The first step, which is called the E step, is to compute the \textit{responsibilities} of cluster $k$ for each datapoint $y_i$:
\begin{align}
  r_t^{i, k} &\coloneqq \frac{\pi_t^k \mathcal{N} (y_i; \mu_t^k, \Sigma_t^k)}{\sum_{k'} \pi_t^{k'} \mathcal{N} (y_i; \mu_t^{k'}, \Sigma_t^{k'})}. \label{resp-01-01}
\end{align}
The second step, which called the M step, is to compute $\theta_t$ by using the responsibilities, Eq.~\eqref{resp-01-01}:
\begin{align}
  \pi_{t+1}^k &= \sum_i r_t^{i, k}, \label{update-EM-GMM-01-01} \\
  \mu_{t+1}^k &= \sum_i \frac{r_t^{i, k} y_i}{r_t^{i, k}}, \label{update-EM-GMM-01-02} \\
  \Sigma_{t+1}^k &= \sum_i \frac{r_t^{i, k} (y_i - \mu_t^k) (y_i - \mu_t^k)^\intercal}{r_t^{i, k}}. \label{update-EM-GMM-01-03}
\end{align}
We iterate the E and M step by substituting Eqs.~\eqref{update-EM-GMM-01-01}, \eqref{update-EM-GMM-01-02}, and \eqref{update-EM-GMM-01-03} until convergence.
Note that we can begin either of the E step or the M steps for the first iteration.
The EM algorithm is summarized in Algo.~\ref{EM-01}.
\begin{algorithm}[t]
\caption{EM algorithm.} \label{EM-01}
\begin{algorithmic}[1]
\STATE $t = 0$
\STATE assign $y_i$ for $i = 1, 2, \dots, N$ to clusters $k = 1, 2, \dots, K$ randomly
\WHILE{convergence criterion is not satisfied}
\STATE compute the responsibilities of cluster $k$ on $r_t^{i, k}$, Eq.~\eqref{resp-01-01}
\STATE estimate $\theta_t = \{ \pi_t^k, \mu_t^k, \Sigma_t^k \}_{k=1}^K$ by Eqs.~\eqref{update-EM-GMM-01-01}, \eqref{update-EM-GMM-01-02}, and \eqref{update-EM-GMM-01-03}
\STATE $t \leftarrow t+1$
\ENDWHILE
\end{algorithmic}
\end{algorithm}
Note that the procedure of the EM algorithm can be generalized for mixture models~\cite{Bishop01, Murphy01}.

\subsection{$\delta$-EM algorithm}

As a prerequisite to the q-EM algorithm, we need to modify the original EM algorithm, since we have to take into account randomness associated with quantum measurement.

In Sec.~\ref{sec-EM-01}, we explained that the EM algorithm has two steps: the E and M steps.
In the $\delta$-EM algorithm, we modify the E step in the spirit of the $\delta$-$k$-means-algorithm in Ref.~\cite{Kerenidis02}.

To this end, we first introduce the square GMM distance by
\begin{align}
d_\mathrm{G}^k (y_i) &\coloneqq  (y_i - \mu^k)^\intercal \Sigma^k (y_i - \mu^k) + \ln | \Sigma^k | - 2 \ln (K \pi^k). \label{dist-GMM-01-01}
\end{align}
Note that, when $\pi^k = 1 / K$ and $\Sigma^k$ is the identity matrix for $k = 1, 2, \dots, K$, $d_\mathrm{G}^k (y_i) = d_\mathrm{E}^k (y_i)$, where $d_\mathrm{E}^k (\cdot)$ is the square Euclidean distance given by
\begin{align}
d_\mathrm{E}^k (y_i) &\coloneqq (y_i - \mu^k)^\intercal (y_i - \mu^k). \label{dist-E-01-01}
\end{align}
We then define the set of labels given by
\begin{align}
L_\mathrm{G}^\delta (y_i) &\coloneqq \Big\{ \mu^k \Big| \big\| d_\mathrm{G}^* (y_i) - d_\mathrm{G}^k (y_i) \big\| \le \delta \Big\}, \label{neighbor-set-GMM-01-01}
\end{align}
where $d_\mathrm{G}^* (y_i) \coloneqq \min_k d_\mathrm{G}^k (y_i)$.
In the E step of the $\delta$-EM algorithm, we take random samples from $L_\mathrm{G}^\delta (y_i)$ in Eq.~\eqref{neighbor-set-GMM-01-01}.
We note that for soft clustering, a more precise sampling scheme may be useful, but this simple sampling works well in the approach shown in Ref.~\cite{Kerenidis02}.

In the M step of the $\delta$-EM algorithm, we add small noise to the estimated parameters after their estimation.
As a result, the $\delta$-EM algorithm becomes robust, and its quantum version will become implementable.
In Sec.~\ref{sec-numerical-simulation}, we will show the validity of this algorithm numerically. Remarkably, we find that adding noise can even improve the quality of the studied benchmark examples.

\section{Quantum algorithm for the EM algorithm with the GMM} \label{quantum-EM-01}

In this section, we describe the procedure of the quantum algorithm that realizes a quantum speedup of the EM algorithm for the GMM.

To simplify the notation, we add the tilde for estimates throughout this paper; that is, we denote e.g. $\tilde{a}$ as the estimate of $a$.

\subsection{Overview of the q-EM algorithm}

We begin with the initialization of the q-EM algorithm.
In the EM algorithm, we can begin either with the E step or the M step for the first iteration.
For simplicity, in the case of the q-EM algorithm, we consider to start with the E step; then, we set the initial parameter set $\theta_0 = \{\pi_0, \mu_0, \Sigma_0\}$ with $\pi_0 \coloneqq [\pi_0^1, \pi_0^2, \dots, \pi_0^K]$, $\mu_0 \coloneqq [\mu_0^1, \mu_0^2, \dots, \mu_0^K]$, and $\Sigma_0 \coloneqq [\Sigma_0^1, \Sigma_0^2, \dots, \Sigma_0^K]$.

The main procedure of the q-EM algorithm is composed of four steps.
In step I, we compute the square GMM distance, and in step II, it is minimized for cluster assignment.
Then, in step III, we generate quantum states of weight vectors, mean vectors, and covariance matrices, and in step IV, we apply quantum vector state tomography.
By using the classical information on mean vectors and covariance matrices, we repeat the whole procedures until convergence.
The output of this algorithm is $\theta_* = \{\pi_*, \mu_*, \Sigma_*\}$.
The q-EM algorithm is summarized in Algo.~\ref{q-EM-01}.
\begin{algorithm}[t]
\caption{q-EM algorithm.} \label{q-EM-01}
\begin{algorithmic}[1]
\STATE $t = 0$
\STATE prepare for the data structures
\WHILE{convergence criterion is not satisfied}
\STATE compute the square GMM distance (step I)
\STATE assign clusters (step II)
\STATE generate the mean and covariance states (step III)
\STATE update the parameters (step IV)
\STATE $t \leftarrow t+1$
\ENDWHILE
\end{algorithmic}
\end{algorithm}
In the rest of this section, we will explain the four steps in detail.

\subsection{Step I: Computing the square GMM distance} \label{sec-q-EM-StepI-02-01}

In this step, we compute the square GMM distance, Eq.~\eqref{dist-GMM-01-01}.
Mathematically, we apply the unitary operation:
\begin{align}
  & \frac{1}{\sqrt{N}} \sum_{i=1}^N \Ket{i} \Big( \otimes_{k \in [K]} \Ket{k} \Ket{0} \Big) \nonumber \\
  & \quad \mapsto \frac{1}{\sqrt{N}} \sum_{i=1}^N \Ket{i} \Big( \otimes_{k \in [K]} \Ket{k} | \tilde{d}_\mathrm{G}^k (y_i) \rangle \Big), \label{StepI-02-01}
\end{align}
where $d_\mathrm{G}^k (y_i)$ is the square GMM distance between $y_i$ and the $k$-th cluster, and $[K] \coloneqq \{ k \}_{k=1}^K$.
For this computation, we require the precision given by $\| \tilde{d}_\mathrm{G}^k (y_i) - d_\mathrm{G}^k (y_i) \| \le \epsilon_1$.
In the next section, $\epsilon_1$ will be used to analyze the runtime.

Equation~\eqref{StepI-02-01} includes summation, multiplication, and inner products of quantum states.
Among them, the computation of summation and multiplication is straightforward with quantum linear algebra while the implementation of inner products is more involved. Let us thus focus on the computation of inner products.

We assume that two unitary operations and their controlled versions are available as follows:
\begin{align}
  \Ket{i} \Ket{0} &\mapsto \Ket{i} \hat{G}^k \Ket{y_i}, \label{assumpution-unitary-01-01} \\
  \Ket{k} \Ket{0} &\mapsto \Ket{k} \hat{G}^k | \mu^k \rangle, \label{assumpution-unitary-01-02}
\end{align}
where $\hat{G}^k \coloneqq (\hat{\Sigma}^k)^{1/2}$ for $k = 1, 2, \dots, K$.
We then begin with the state
\begin{align}
  \Ket{\phi_{i, k}} &\coloneqq \Ket{i} \Ket{k} \frac{1}{\sqrt{2}} (\Ket{0} + \Ket{1}) \Ket{0}. \label{phi-start-01-01}
\end{align}
By using controlled versions of Eq.~\eqref{assumpution-unitary-01-01} and \eqref{assumpution-unitary-01-02}, we create the following state from $\Ket{\phi_{i, k}}$:
\begin{align}
  | \phi_{i, k}^C \rangle &\coloneqq \frac{1}{\sqrt{2}} (\Ket{i} \Ket{k} \Ket{0} \hat{G}^k \Ket{y_i} + \Ket{i} \Ket{k} \Ket{1} \hat{G}^k | \mu^k \rangle).
\end{align}
Then we apply the Hadamard gate on the third register of $| \phi_{i, k}^C \rangle$ and the resulting state is
\begin{align}
  | \phi_{i, k}^H \rangle = \frac{1}{2} \Ket{i} \Ket{k} \Big(& \Ket{0} (\hat{G}^k \Ket{y_i} + \hat{G}^k | \mu^k \rangle) \nonumber \\
  & + \Ket{1} (\hat{G}^k \Ket{y_i} - \hat{G}^k | \mu^k \rangle) \Big). \label{state-H-02-01}
\end{align}
Note that Eq.~\eqref{state-H-02-01} is also represented as
\begin{align}
  | \phi_{i, k}^H \rangle &= \Ket{i} \Ket{k} \Big( \sqrt{p_{i, k}} \Ket{\mathrm{tar}_{i, k}, 1} + \sqrt{1 - p_{i, k}} \Ket{\mathrm{gar}_{i, k}, 0} \Big), \label{state-H-02-02}
\end{align}
where $\Ket{\mathrm{tar}_{i, k}, 1} \coloneqq \Ket{1} \hat{G}^k (\Ket{y_i} - | \mu^k \rangle)$ and $\Ket{\mathrm{gar}_{i, k}, 0}$ is a garbage state.
That is, we have a unitary operator such that
\begin{align}
  \hat{U}_1 &: \Ket{i} \Ket{j} \Ket{0} \nonumber \\
  & \quad \mapsto \Ket{i} \Ket{j} \Big( \sqrt{p_{i, k}} \Ket{\mathrm{tar}_{i, k}, 1} + \sqrt{1 - p_{i, k}} \Ket{\mathrm{gar}_{i, k}, 0} \Big). \label{unitary-operator-01-01}
\end{align}
We also note that the probability that we get $\Ket{1}$ by measuring the third register is expressed as
\begin{align}
  p_{i, k} = \frac{1 - \langle y_i| (\hat{G}^k)^2| \mu^k \rangle}{2}.
\end{align}

We then apply all the operations except the measurement in amplitude estimation in Ref.~\cite{Brassard01, Kerenidis02} on $\hat{U}_1$ in Eq.~\eqref{unitary-operator-01-01}.
This process realizes the following unitary operation:
\begin{align}
  \hat{U}_2 &: \Ket{i} \Ket{j} \Ket{0} \nonumber \\
  & \quad \mapsto \Ket{i} \Ket{k} (\sqrt{\alpha} \Ket{\tilde{p}_{i, k}, 1} + \sqrt{1 - \alpha} \Ket{\mathrm{gar}_{i, k}, 0}), \label{state-stepI-03-01}
\end{align}
where $\| \tilde{p}_{i, k} - p_{i, k} \| < 2 \pi \sqrt{p_{i, k} (1 - p_{i, k})} / P_\mathrm{ae} + \pi^2 / P_\mathrm{ae}^2$ and $\alpha > \frac{8}{\pi^2}$~\cite{Brassard01}.
Here $M$ is a parameter to be determined (see Sec.~\ref{amp-est-01-01}).
Next, applying the mode evaluation method~\footnote{This algorithm has no specific name in Ref.~\cite{Wiebe01}, and it is called median evaluation in Ref.~\cite{Kerenidis02}. But this algorithm realize majority voting; so we call it mode evaluation.} in Lemma~8 of Ref.~\cite{Wiebe01} and Thm.~2.2 of \cite{Kerenidis02} to Eq.~\eqref{state-stepI-03-01}, we get $\Ket{\Phi_{i, k}}$ such that
\begin{align}
  \| \Ket{\Phi_{i, k}} - \Ket{0}^{\otimes L} \Ket{\tilde{p}_{i, k}} \|_2 &\le \sqrt{2 \Delta}.
\end{align}

The last step is to estimate the square GMM distance of unnormalized vectors $\| y_i \|$ and $\| \mu^k \|$ and to multiply the norms of them and adding $\ln \pi^k$.
A translation operator $\hat{T} (r')$ can conduct the adding operation: $\hat{T} (r') \Ket{r} = \Ket{r + r'}$ for $r, r' \in \mathbb{R}^N$.
Note that we have assumed that we know the norms of $\{ \| y_i \| \}_{i=1}^N$ and the same assumption is used in Ref.~\cite{Kerenidis02}.

\subsection{Step II: Assignment of clusters}

The purpose of step II is cluster assignment.
In this step, we utilize the following unitary operation:
\begin{align}
  \hat{U}_3 &: \Big( \otimes_{k \in [K]} | a^k \rangle \Big) \Ket{0} \mapsto \Big( \otimes_{k \in [K]} | a^k \rangle \Big) |\argmin_{k \in [K]} a^k \rangle,
\end{align}
where $| a^k \rangle$ is a $(\ln p)$-bit state for $k = 1, 2, \dots, K$.
The computational cost of this operation is $O (K \ln p)$~\cite{Kerenidis02}.

To find cluster assignment, we perform the unitary operation given by
\begin{align}
  \hat{U}_4 &: \frac{1}{\sqrt{N}} \sum_{i=1}^N \Ket{i} \Big( \otimes_{k \in [K]} \Ket{k} | d_\mathrm{G}^k (y_i) \rangle \Big) \nonumber \\
  & \quad \mapsto \frac{1}{\sqrt{N}} \sum_{i=1}^N \Ket{i} \Ket{\mathrm{label}_t (y_i)}, \label{unitary-stepII-01-01}
\end{align}
where $\mathrm{label}_t (y_i)$ is the optimal label of $y_i$ at time $t$.

Finally, by uncomputing the square GMM distances, we obtain
\begin{align}
  | \psi_t \rangle &\coloneqq \frac{1}{\sqrt{N}} \sum_{i=1}^N \Ket{i} \Ket{\mathrm{label}_t (y_i)}. \label{psi-01-01}
\end{align}
This uncomputation is required to repeat iterations.

\subsection{Step III: Generation of the mean and covariance states}

In this step, we generate states that store information on the weights, mean vectors, and covariance matrices.
Let us recall that Eq.~\eqref{psi-01-01} is also expressed as
\begin{align}
  \Ket{\psi_t} &= \sum_{k=1}^K \sqrt{\frac{N_t^k}{N}} \Bigg( \frac{1}{\sqrt{N_t^k}} \sum_{i \in C_t^k} \Ket{i} \Bigg) \Ket{k}  = \\
  &= \sum_{k=1}^K \sqrt{\frac{N_t^k}{N}} | \chi_t^k \rangle \Ket{k}.
\end{align}
Thus, by measuring the label register of $| \psi_t \rangle$ in Eq.~\eqref{psi-01-01}, we obtain, with probability $N_t^k / N$,
\begin{align}
  | \chi_t^k \rangle &= \frac{1}{\sqrt{N_t^k}} \sum_{i \in C_t^k} \Ket{i}, \label{chi-01-01}
\end{align}
where $C_t^k$ is the set of labels that belong to cluster $k$ at time $t$.
Then, $\chi_t^k = [\dots, (\chi_t^k)_{i-1}, (\chi_t^k)_i, (\chi_t^k)_{i+1}, \dots]^\intercal \in \mathbb{R}^N$ is
\begin{align}
  (\chi_t^k)_i &=
  \begin{cases}
    1 / N_t^k & (i \in C_t^k), \\
    0 & (i \not\in C_t^k),
  \end{cases}
\end{align}
for $i = 1, 2, \dots, N$.

We here define $V_1 \in \mathbb{R}^{N \times d}$, $V_2 \in \mathbb{R}^{N \times d \times d}$, and $V_{0, i}$ for $i = 1, 2, \dots, N$ on a QRAM:
\begin{align}
  V_1 &\coloneqq [y_1, y_2, \dots, y_N], \label{data-matrix-02-01} \\
  V_2 &\coloneqq [y_1 \otimes y_1, y_2 \otimes y_2, \dots, y_N \otimes y_N], \label{data-matrix-02-02} \\
  V_{0, i} &\coloneqq [\underbrace{\vec{0}, \vec{0}, \dots, \vec{0}}_{i-1}, \vec{1}, \underbrace{\vec{0}, \vec{0}, \dots, \vec{0}}_{N - i - 1}]. \label{data-matrix-02-03}
\end{align}

To obtain mean vectors, we multiply $V_1$ to the state $| \chi_t^k \rangle$ in Eq.~\eqref{chi-01-01} by using quantum linear algebra~\cite{Chakraborty01, Kerenidis03}:
\begin{align}
  | \mu_{t+1}^k \rangle &= V_1 | \chi_t^k \rangle.
\end{align}
The associated error is $\epsilon_2^\mu$.
Similarly, we compute a state involving information on $\Sigma^k$ by using quantum linear algebra~\cite{Chakraborty01, Kerenidis03}:
\begin{align}
  | \mathrm{vec} [\Sigma_{t+1}^k] + \mu_{t+1}^k \otimes \mu_{t+1}^k \rangle &= V_2 | \chi_t^k \rangle.
\end{align}
The associated error is $\epsilon_2^\Sigma$.
Note that $\epsilon_2^\mu$ and $\epsilon_2^\Sigma$ appear only in logarithms; thus, we do not explicitly consider them.

We finally deal with $\{ \pi^k \}_{k=1}^K$; it is relatively easy to compute the weights of the GMM, $\{\pi_t^k\}_{k=1}^K$.
We here utilize Eq.~\eqref{data-matrix-02-03} as follows:
\begin{align}
(V_{0, i}) \chi^k &=
\begin{cases}
(N^k)^{-1} \vec{1} & (i \in C_t^k), \\
\vec{0} & (\mathrm{otherwise}).
\end{cases}
\end{align}
Thus, we can estimate $N_t^k$ similarly.
Note that we assume that the sizes of all clusters are $\Omega (N / k)$.

\subsection{Step IV: Update of the parameters}

At the end of each iteration, we obtain classical information on $\pi_{t+1}$, $\mu_{t+1}$, and $\Sigma_{t+1}$ by performing the quantum state tomography algorithm for $| \chi_t^k \rangle$, $|\mu_{t+1}^k \rangle$, and $| \mathrm{vec} [\Sigma_{t+1}^k] \rangle$.
Quantum vector state tomography is explained in Ref.~\cite{Kerenidis02}.
The quantum state tomography algorithm in Ref.~\cite{Kerenidis02} requires a unitary transformation $U: \Ket{0} \mapsto \Ket{x}$; however, the procedure to find $l (y_i)$ is not deterministic.
Then, we have to devise some deterministic methods to find $l (y_i)$.
One solution is to determine $l (y_i)$ by the rule
\begin{align}
  l (y_i) &= k,
\end{align}
if $d_\mathrm{G}^k (y_i) < d_\mathrm{G}^{k'} (y_i) - 2 \delta$ for $k' \ne k$, and we discard the points to which no label can be assigned.

By introducing $\epsilon_4^\pi$, $\epsilon_3^\mu$, $\epsilon_4^\mu$, $\epsilon_3^\Sigma$, and $\epsilon_4^\Sigma$, we require the following precision in this step: $\big\| \| \pi^k \|- \| \tilde{\pi}^k \| \big\| \le \epsilon_4^\pi$, $\big\| | \mu^k \rangle - | \tilde{\mu}^k \rangle \big\| \le \epsilon_3^\mu$, $\big\| \| \mu^k \|- \| \tilde{\mu}^k \| \big\| \le \epsilon_4^\mu \| \mu^k \|$, $\big\| | \Sigma^k \rangle - | \tilde{\Sigma}^k \rangle \big\| \le \epsilon_3^\mu$, and $\big\| \| \Sigma^k \|- \| \tilde{\Sigma}^k \| \big\| \le \epsilon_4^\mu \| \Sigma^k \|$.
In the next section, $\epsilon_4^\pi$, $\epsilon_3^\mu$, $\epsilon_4^\mu$, $\epsilon_3^\Sigma$, and $\epsilon_4^\Sigma$ will be used to analyze the runtime.


\section{Analysis of errors and runtime} \label{sec-analysis-02-01}

This section is dedicated to error and runtime analysis of the q-EM algorithm.
We first state the main claim and then explain it.

\subsection{Main result}

The runtime of the q-EM algorithm is represented by
\begin{widetext}
\begin{align}
  \tilde{O} \bigg(\frac{K^2}{\epsilon_1 (\epsilon_4^\pi)^2} + K d \frac{\kappa (V_1)}{(\epsilon_4^\mu)^2} \bigg( \mu (V_1) + K \frac{\eta^\mu}{\epsilon_1} \bigg) + \frac{K^2}{\epsilon_1} \frac{\eta^\mu \kappa (V_1) \mu (V_1)}{\epsilon_3^\mu} + K d^2 \frac{\kappa (V_2)}{(\epsilon_4^\Sigma)^2} \bigg( \mu (V_2) + K \frac{\eta^\Sigma}{\epsilon_1} \bigg) + \frac{K^2}{\epsilon_1} \frac{\eta^\Sigma \kappa (V_2) \mu (V_2)}{\epsilon_3^\Sigma} \bigg), \label{runtime-qEM-01-01}
\end{align}
\end{widetext}
where $\mu (\cdot)$ is given in Eq.~\eqref{def-mu-01-01}, $\kappa (\cdot)$ is the condition number, $\eta^\mu \coloneqq \max_i \| y_i \|^2$, and $\eta^\Sigma \coloneqq \max_i \| y_i \otimes y_i \|^2$.
The definition of $\tilde{O} (\cdot)$ is given in Appendix~\ref{sec-big-o-notation-01}.

This result states that the runtime of each iteration of the q-EM algorithm is exponentially faster than that of the EM algorithm.

\subsection{Error analysis}

We first summarize the errors in the q-EM algorithm to analyze the total runtime of the q-EM algorithm in the following subsection.
In step I, we compute $d_\mathrm{G}^k (y_i)$; the error on this computation is
\begin{align}
  \| \tilde{d}_\mathrm{G}^k (y_i) - d_\mathrm{G}^k (y_i) \| &< \epsilon_1.
\end{align}
For consistency between the EM algorithm and the $\delta$-EM algorithm, we take $\epsilon_1 < \delta / 2$.

In steps III and IV, we compute $\mu$ and $\Sigma$.
The errors on $\| \mu^k \|$ and $| \mu^k \rangle$ are $\sqrt{\eta^\mu} \epsilon_3^\mu$ and $\epsilon_4^\mu$, respectively.
Then, the error on the estimation of $\mu$ takes the form
\begin{align}
  \| \tilde{\mu}^k - \mu^k \| &\le \sqrt{\eta^\mu} (\epsilon_3^\mu + \epsilon_4^\mu). \label{error-bound-01-31}
\end{align}
See also Appendix~\ref{sec-supplementary-calculation} for the above calculation.
We need to take $\epsilon_3^\mu < \frac{\delta}{4 \sqrt{\eta^\mu}}$ and $\epsilon_4^\mu < \frac{\delta}{4 \sqrt{\eta^\mu}}$.
Next, we turn our attention to $\Sigma$.
The errors on $\sqrt{\eta^\Sigma} \epsilon_3^\Sigma$ and $\sqrt{\eta^\Sigma} \epsilon_4^\Sigma$ are $\sqrt{\eta^\Sigma} \epsilon_3^\Sigma$ and $\epsilon_4^\Sigma$, respectively.
Similarly to the case of $\mu^k$, the error on the estimation of $\Sigma$ is shown as
\begin{align}
  \| \mathrm{vec} [\tilde{\Sigma}^k] - \mathrm{vec} [\Sigma^k] \| &\le \sqrt{\eta^\Sigma} (\epsilon_3^\Sigma + \epsilon_4^\Sigma).
\end{align}
We also need to take $\epsilon_3^\Sigma < \frac{\delta}{4 \sqrt{\eta^\Sigma}}$ and $\epsilon_4^\Sigma < \frac{\delta}{4 \sqrt{\eta^\Sigma}}$.
Finally, we mention the error associated with the estimation on $\pi^k$ for $k = 1, 2, \dots, K$.
The error on $\pi^k$ is shown as
\begin{align}
\| \tilde{\pi}^k - \pi^k \| &\le \epsilon_4^\pi.
\end{align}
We estimate $\{ \pi^k \}_{k=1}^K$ via the distribution of labels in quantum vector state tomography.

\subsection{Runtime}

In the following, Eq.~\eqref{runtime-qEM-01-01}, i.e the runtime of each iteration of the q-EM algorithm is derived using Hoeffding's inequality
The required number of quantum vector state tomography of $K$ mean vectors is given as follows~\cite{Kerenidis02}:
\begin{align}
  \tilde{O} \bigg( \frac{K d \ln K \ln d}{(\epsilon_4^\mu)^2} \bigg).
\end{align}
Similarly, that for covariance matrices is
\begin{align}
  \tilde{O} \bigg( \frac{K d^2 \ln K \ln d^2}{(\epsilon_4^\Sigma)^2} \bigg).
\end{align}
The definition of $\tilde{O} (\cdot)$ is given in Appendix~\ref{sec-big-o-notation-01}.
Next, we turn our attention to the runtime to prepare single copies of $| \mu^k \rangle$ and $| \mathrm{vec} [\Sigma^k] \rangle$.
The time to prepare a copy of $| \mu_t^k \rangle$ is
\begin{align}
  O \Big( \kappa (V_1) (\mu (V_1) + T_\chi^\mu) \ln (1 / \epsilon_2) \Big),
\end{align}
and that to prepare a copy of $| \mathrm{vec} [\Sigma_t^k] \rangle$ is
\begin{align}
  O \Big( \kappa (V_2) (\mu(V_2) + T_\chi^\Sigma) \ln (1 / \epsilon_2) \Big),
\end{align}
where $\mu (V_1)$ is given in Eq.~\eqref{def-mu-01-01}, $\kappa (V_1)$ is the condition number of $V_1$, $\eta^\mu \coloneqq \max_i \| y_i \|^2$ and $\eta^\Sigma \coloneqq \max_i \| y_i \otimes y_i \|^2$.
Furthermore, $T_\chi^\mu$, which is the time to prepare $| \chi_t^k \rangle$ for estimating $\{ \mu^k \}_{k=1}^K$, is given by
\begin{align}
  T_\chi^\mu &= \tilde{O} \bigg( \frac{K \eta^\mu \ln (\Delta^{-1}) \ln (N d)}{\epsilon_1} \bigg) \\
  &= \tilde{O} \bigg( \frac{K \eta^\mu}{\epsilon_1} \bigg).
\end{align}
Similarly, $T_\chi^\Sigma$ is given by
\begin{align}
  T_\chi^\Sigma &= \tilde{O} \bigg( \frac{K \eta^\Sigma \ln (\Delta^{-1}) \ln (N d)}{\epsilon_1} \bigg) \\
  &= \tilde{O} \bigg( \frac{K \eta^\Sigma}{\epsilon_1} \bigg).
\end{align}
In addition, $T_\chi^\pi$ is given by
\begin{align}
  T_\chi^\pi &= \tilde{O} \bigg( \frac{K  \ln (\Delta^{-1}) \ln (N d)}{\epsilon_1} \bigg) \\
  &= \tilde{O} \bigg( \frac{K}{\epsilon_1} \bigg).
\end{align}

We also need to estimate the norms of $| \mu^k \rangle$ and $| \mathrm{vec} [\Sigma^k] \rangle$.
The time for the norm estimation of $| \mu^k \rangle$ is
\begin{align}
  \tilde{O} \bigg( \frac{K T_\chi^\mu \kappa (V_1) \mu (V_1)}{\epsilon_3^\mu} \bigg),
\end{align}
and that of $| \mathrm{vec} [\Sigma^k] \rangle$ is
\begin{align}
  \tilde{O} \bigg( \frac{K T_\chi^\Sigma \kappa (V_2) \mu (V_2)}{\epsilon_3^\Sigma} \bigg).
\end{align}

We then estimate the runtime for estimating $\{ \pi^k \}_{k=1}^K$.
Due to Hoeffding's inequality~\cite{Hoeffding01}, we need to perform sampling $\frac{2K}{(\epsilon_4^\pi)^2} \ln \frac{2}{\Delta_\mathrm{each}^\pi}$ times to realize $\| \tilde{p}_k - p_k \| \le \epsilon_4^\pi$ for $k = 1, 2, \dots, K$ with probability $(1 - \Delta_\mathrm{each}^\pi)^K$ since the distribution of $\pi$ is the $K$-state discrete distribution.
By setting $1 - \Delta^\pi \coloneqq (1 - \Delta_\mathrm{each}^\pi)^K$, we have $\Delta_\mathrm{each}^\pi = 1 - (1 - \Delta^\pi)^{1 / K}$.
Thus, we have
\begin{align}
N_\pi &= \frac{2K}{(\epsilon_4^\pi)^2} \ln \frac{2}{1 - (1 - \Delta^\pi)^{1 / K}} \\
&= \tilde{O} \bigg( \frac{K}{(\epsilon_4^\pi)^2} \bigg).
\end{align}
Furthermore, we have to repeat the estimation process $K$ times for estimation of $\pi$ compared to those of $\mu$ and $\Sigma$, since we have to sample $i$ from $C^k$, $K$ times, for $k = 1, 2, \dots, K$.
Thus, the runtime for estimating $\pi$ has an additional multiple of $K$.

Thus, the total runtimes for estimating $\{ \pi^k \}$, $\{ \mu^k \}$, and $\{ \Sigma^k \}$ are, respectively,
\begin{align}
  \tilde{O}& \bigg(  \frac{K^3}{\epsilon_1 (\epsilon_4^\pi)^2} \bigg), \\
  \tilde{O}& \bigg( K d \frac{\kappa (V_1)}{(\epsilon_4^\mu)^2} \bigg( \mu (V_1) + K \frac{\eta^\mu}{\epsilon_1} \bigg) + \frac{K^2}{\epsilon_1} \frac{\eta^\mu \kappa (V_1) \mu (V_1)}{\epsilon_3^\mu} \bigg), \\
  \tilde{O}& \bigg( K d^2 \frac{\kappa (V_2)}{(\epsilon_4^\Sigma)^2} \bigg( \mu (V_2) + K \frac{\eta^\Sigma}{\epsilon_1} \bigg) + \frac{K^2}{\epsilon_1} \frac{\eta^\Sigma \kappa (V_2) \mu (V_2)}{\epsilon_3^\Sigma} \bigg).
\end{align}
In total, we have obtained Eq.~\eqref{runtime-qEM-01-01}.

\section{Numerical simulation} \label{sec-numerical-simulation}

To devise a quantum version of the EM algorithm, we proposed the $\delta$-EM algorithm in Sec.~\ref{classical-preliminaries-01}.
In this section, to see that the EM and $\delta$-EM algorithms are equivalent when $\delta$ is sufficiently small and that the $\delta$-EM algorithm improves upon the $\delta$-$k$-means algorithm, we show numerical simulations of the EM algorithm, the $\delta$-EM algorithm, the $k$-means algorithm, and the $\delta$-$k$-means algorithm.

In Ref.~\cite{Cornell01}, the comparison between the $k$-means algorithm and the EM algorithm with GMM is shown.
Then we use similar synthetic data sets used in Ref.~\cite{Cornell01}.
In the numerical simulations, we set $\delta = 0.2$ except the numerical simulation for the $\delta$-dependence of the $\delta$-EM algorithm.
For simplicity, we add Gaussian noise to the parameters estimated in the M step of the $\delta$-EM algorithm and the centroids estimated in the $\delta$-$k$-means algorithm~\footnote{In the the $\delta$-$k$-means algorithm, we added Gaussian noise whose mean is 0 and variance is 0.01 to each element of all the centroids. In the M step of the $\delta$-EM algorithm, we added Gaussian noise whose mean is 0 and variance is 0.01 to $\pi^k$ and each element of $\mu^k$, and add Gaussian noise whose mean is 0 and variance is 0.001 to each element of $\Sigma^k$. When $\pi$ is not normalized, we normalized $\pi$. When $\Sigma^k$ is not symmetric, we symmetrize $\Sigma^k$ by $(\Sigma^k + (\Sigma^k)^\intercal) / 2$. When the estimated $\Sigma^k$ has a negative eigenvalue $\sigma^*$, we make all eigenvalues positive by adding $| \sigma^* | I^k$ where $I^k$ is the $k \times k$ identity matrix}.

\subsection{Example I}

We begin with the explanation of the data set used in this subsection.
We generated by drawing 1000 data points from the mixture of two Gaussian functions.
The means of the two Gaussian functions are $\mu^1 = [0.3, 0.0]^\intercal$ and $\mu^2 = [-0.3, 0.0]^\intercal$, respectively, and the covariances are, respectively,
\begin{align}
  \Sigma^1 &=
  \begin{bmatrix}
    1.0 & 0.98 \\
    0.98 & 1.0
  \end{bmatrix}, \\
  \Sigma^2 &=
  \begin{bmatrix}
    1.0 & -0.98 \\
    -0.98 & 1.0
  \end{bmatrix}.
\end{align}
We also put $\pi^1 = \pi^2 = 0.5$.

\begin{figure}[t]
\begin{subfigure}[t]{0.22\textwidth}
\centering
\includegraphics[scale=0.25]{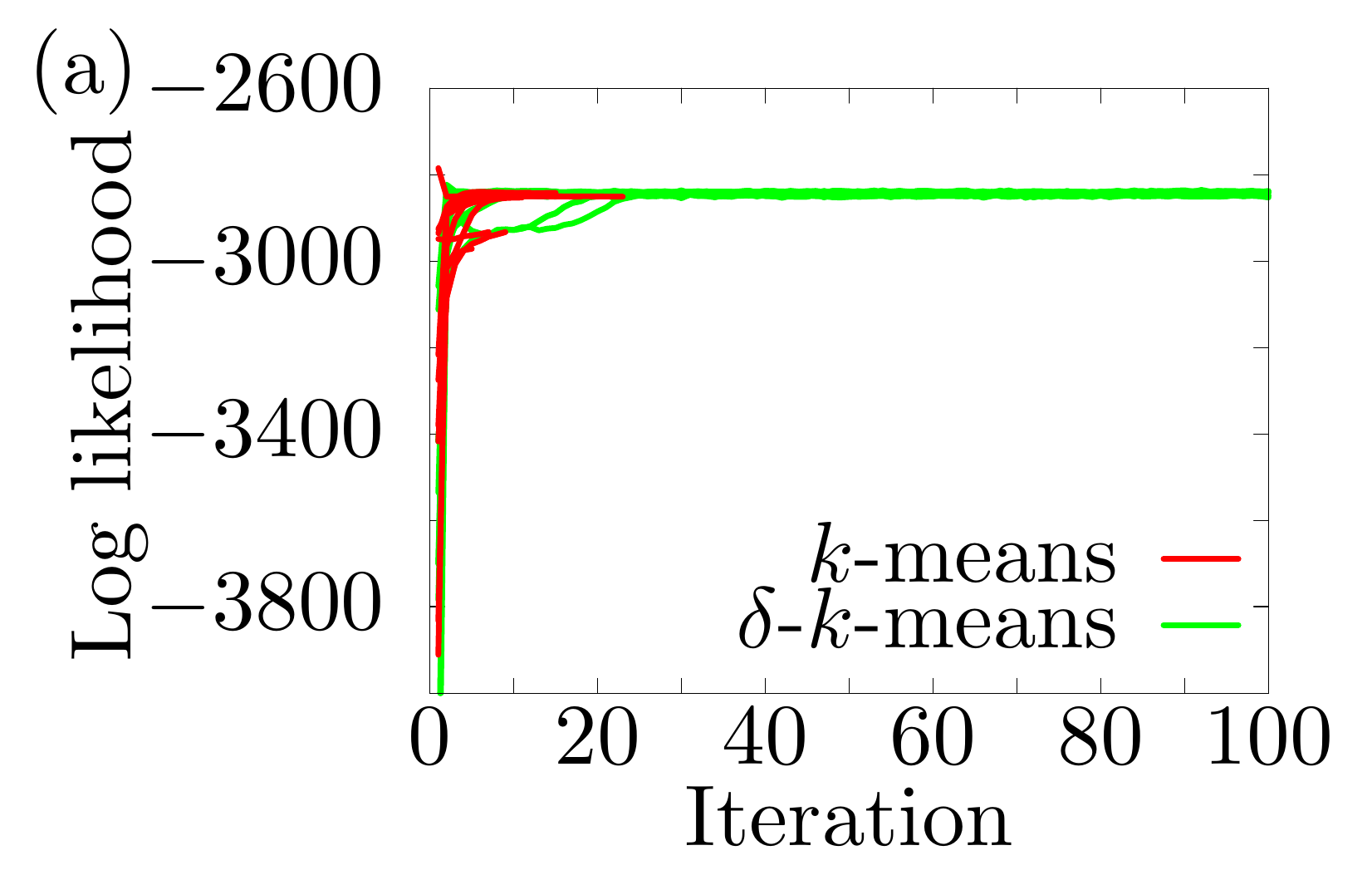}
\end{subfigure}
\begin{subfigure}[t]{0.22\textwidth}
\centering
\includegraphics[scale=0.25]{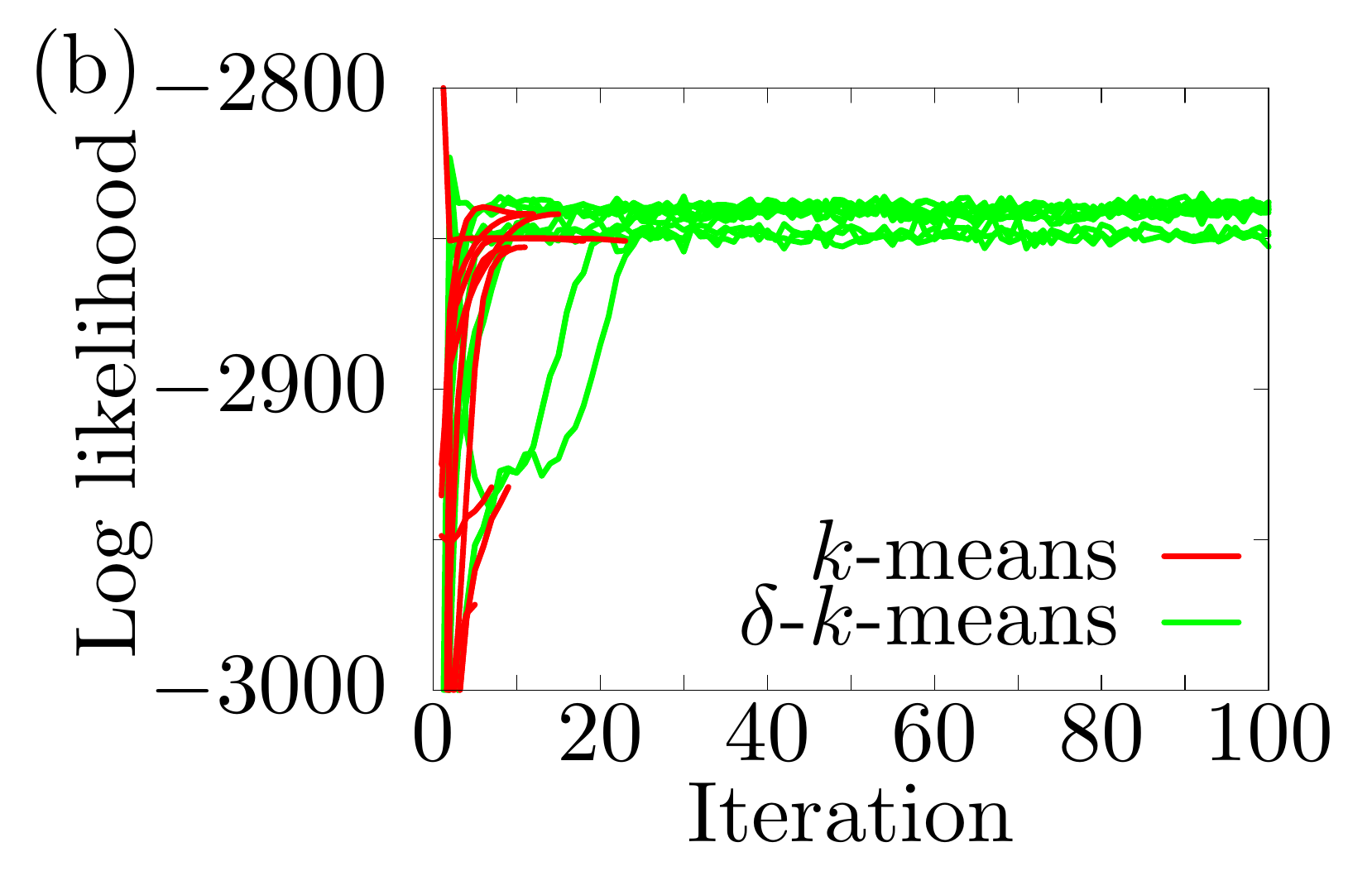}
\end{subfigure}
\caption{Log-likelihood ofthe $k$-means algorithm  (red lines) and the $\delta$-$k$-means algorithm (green lines). We perform the simulation ten times, respectively. The difference of (a) and (b) is the scale of the vertical axis.}
\label{numerical-01-02}
\end{figure}

\begin{figure}[t]
\begin{subfigure}[t]{0.22\textwidth}
\centering
\includegraphics[scale=0.25]{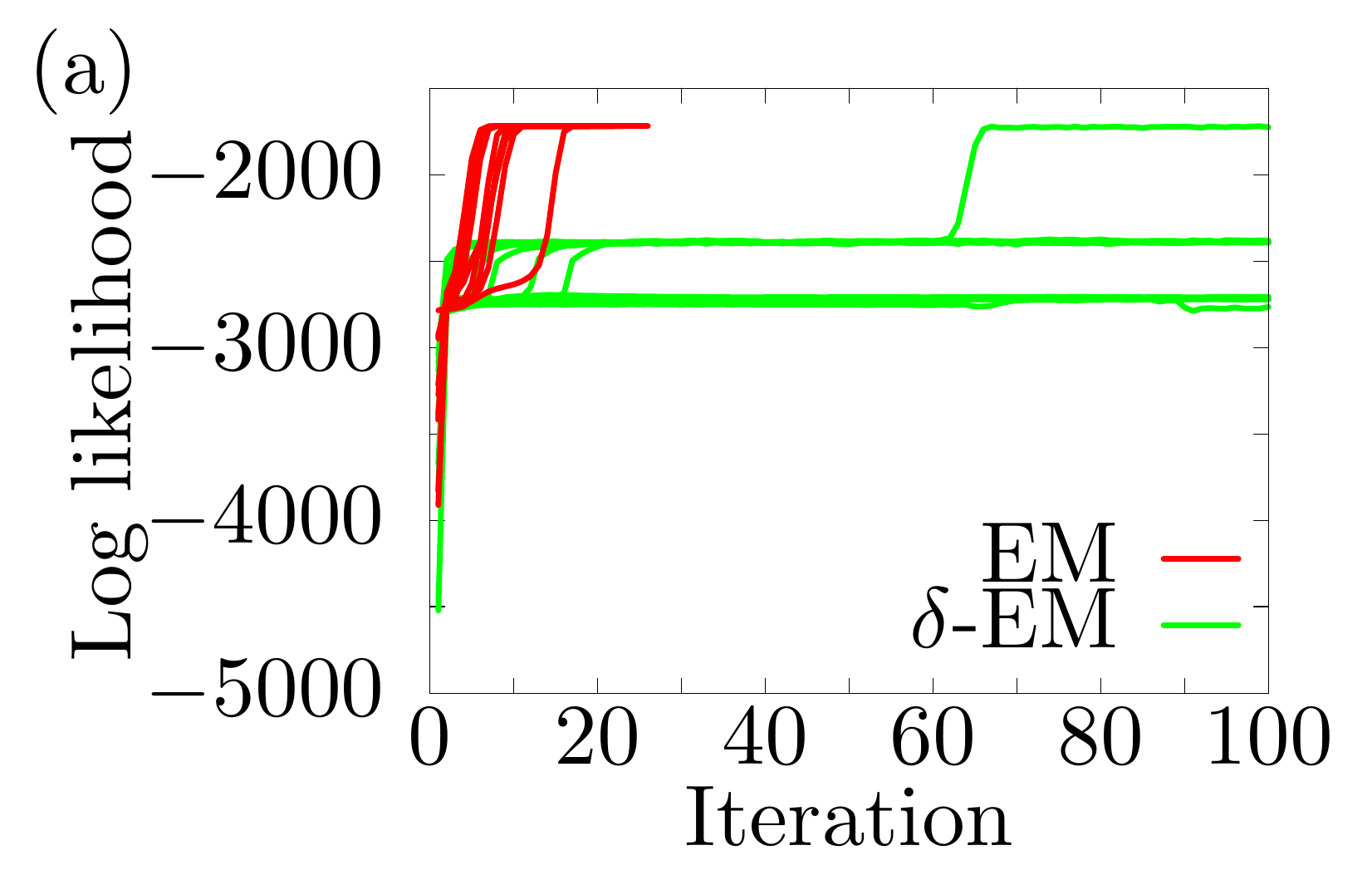}
\end{subfigure}
\begin{subfigure}[t]{0.22\textwidth}
\centering
\includegraphics[scale=0.25]{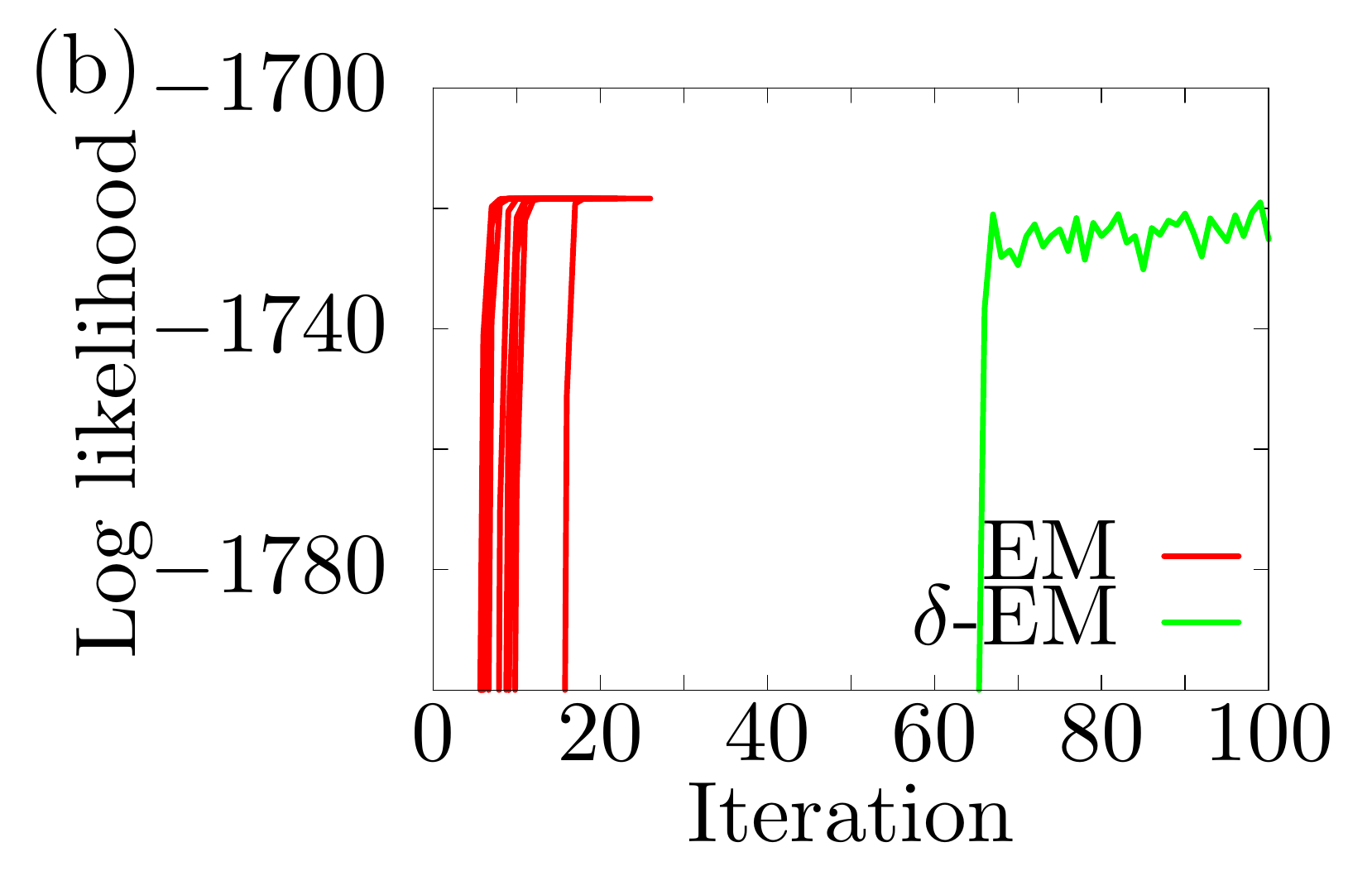}
\end{subfigure}
\caption{Log-likelihood of the EM algorithm (red lines) and the $\delta$-EM algorithm (green lines). We perform the simulation ten times, respectively. The difference of (a) and (b) is the scale of the vertical axis.}
\label{numerical-01-03}
\end{figure}

\begin{figure}[t]
\begin{subfigure}[t]{0.22\textwidth}
\centering
\includegraphics[scale=0.30]{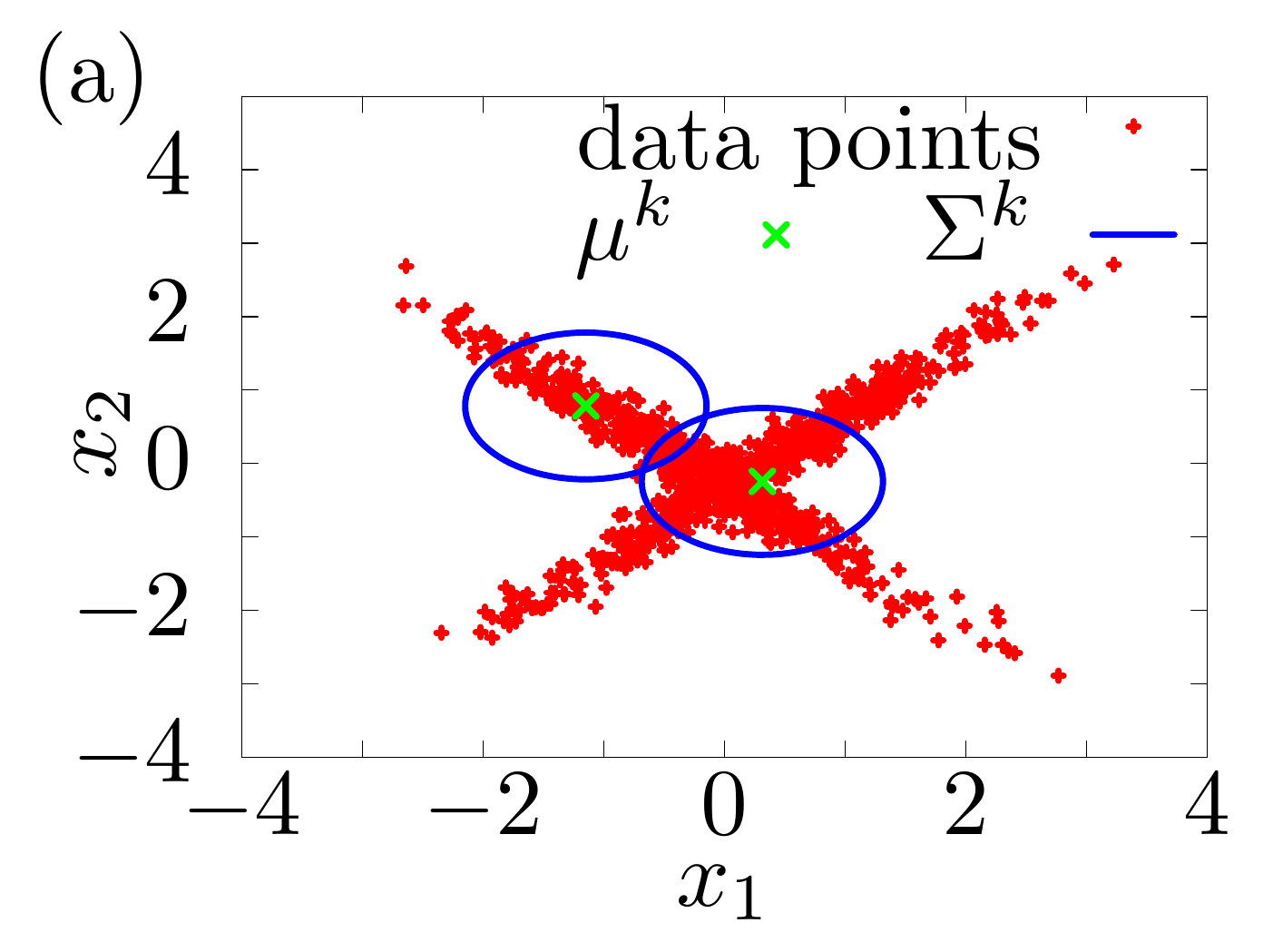}
\end{subfigure}
\begin{subfigure}[t]{0.22\textwidth}
\centering
\includegraphics[scale=0.30]{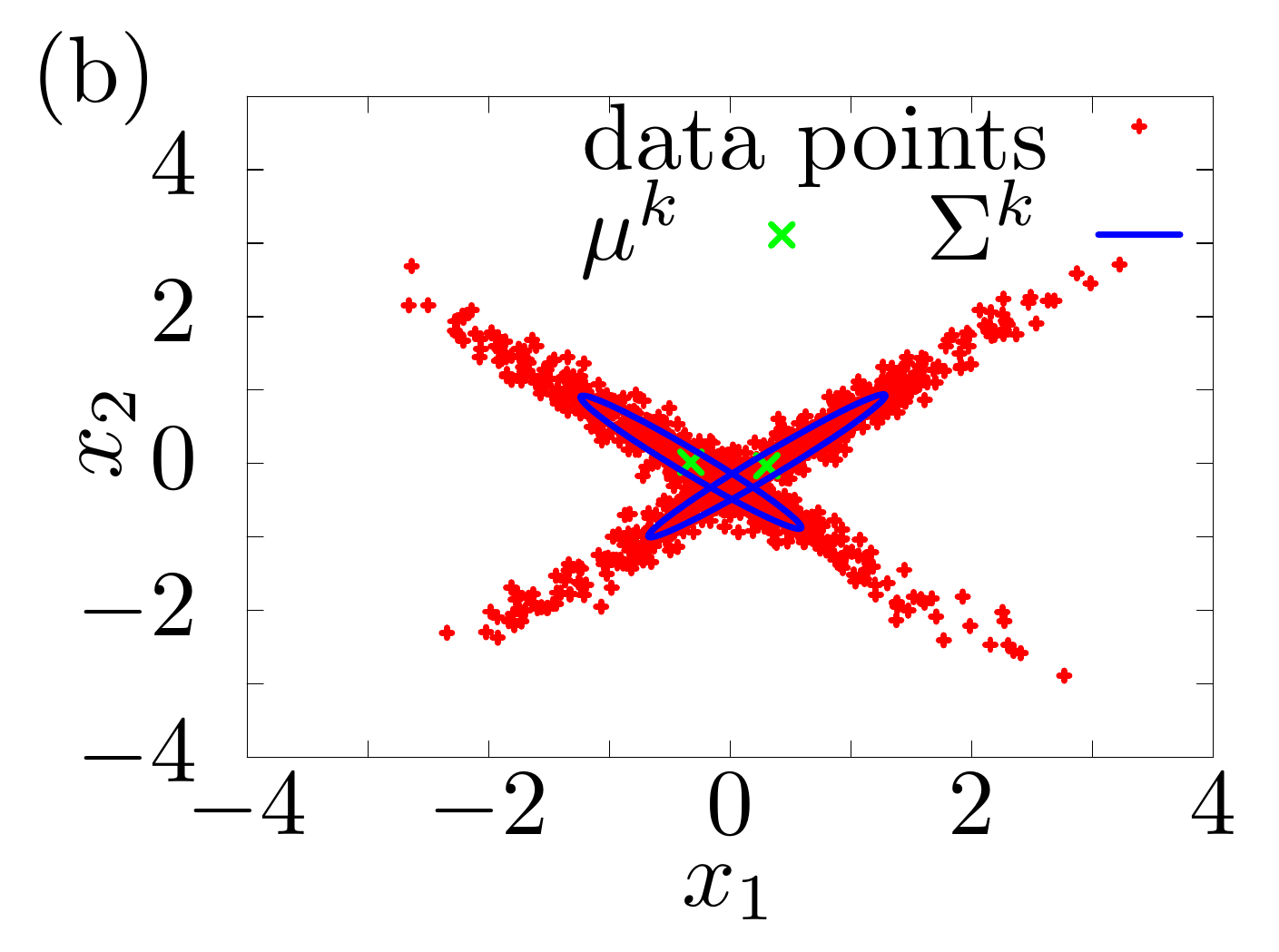}
\end{subfigure}
\caption{Pictures of estimated functions by (a)~the $\delta$-$k$-means algorithm and (b)~the $\delta$-EM algorithm.}
\label{numerical-01-04}
\end{figure}

In Figs.~\ref{numerical-01-02} and \ref{numerical-01-03}, we show the log-likelihood of the $k$-means algorithm and the $\delta$-$k$-means algorithm, and that of the EM algorithm and the $\delta$-EM algorithm, respectively.
We plot ten trials for each algorithm.
These figures show that the $k$-means and $\delta$-$k$-means algorithms have similar performance and that the EM and $\delta$-EM algorithms have similar performance.
For clarity, we graphically show the parameters estimated by the $\delta$-$k$-means algorithm and the $\delta$-EM algorithm in Fig.~\ref{numerical-01-04}~\footnote{In Fig.~\ref{numerical-01-04}, we draw the blue ellipses by moving $\theta$ $(0 \le \theta < 2 \pi)$ in $f (\theta) \coloneqq \mu^k + \sqrt{\sigma_1^k} e_1^k \cos (\theta) + \sqrt{\sigma_2^k} e_2^k \sin \theta$ where $\sigma_1^k$ and $\sigma_2^k$ are the eigenvalues of $\Sigma^k$, and $e_1^k$ and $e_2^k$ are the corresponding eigenvectors of $\Sigma^k$, respectively, for $k = 1, 2$. In Fig.~\ref{numerical-02-04}, we also draw blue ellipses by the same procedure.}.
We have chosen the best estimates of the $\delta$-$k$-means algorithm and the $\delta$-EM algorithm in one hundred trials.
These figures demonstrate that the $\delta$-EM algorithm outperforms the $\delta$-$k$-means algorithm.

In Table~\ref{success-rate-01-01}, we summarize the success rates of the EM algorithm, the $\delta$-EM algorithm, the $k$-means algorithm, and the $\delta$-$k$-means algorithm.
Here, the success rate means that the ratio of the number of the successfully predicted hidden variables~\footnote{The hidden variable in the GMM means the index of the Gaussian functions. Note that the GMM has the symmetry on swapping the indices; then, we use the maximum value on swapping.} to the number of the total data points.
This table shows that the $\delta$-EM algorithm works better than the $\delta$-$k$-means algorithm.
Thus, we insist that it is meaningful to devise a quantum version of the $\delta$-EM algorithm.
\begin{table}
\centering
\begin{tabular}{c c c c}
  \hline
  \hline
  EM & $\delta$-EM  & $k$-means & $\delta$-$k$-means \\
  \hline
  93.9 \% & 94.3 \%  & 72.4 \% & 72.5 \% \\
  \hline
  \hline
\end{tabular}
\caption{Success rates of the EM algorithm, the $\delta$-EM algorithm, the $k$-means algorithm, and the $\delta$-$k$-means algorithm. These scores are best ones in one hundred trials with randomized initial inputs.}
\label{success-rate-01-01}
\end{table}

\begin{figure}[t]
\centering
\includegraphics[scale=0.45]{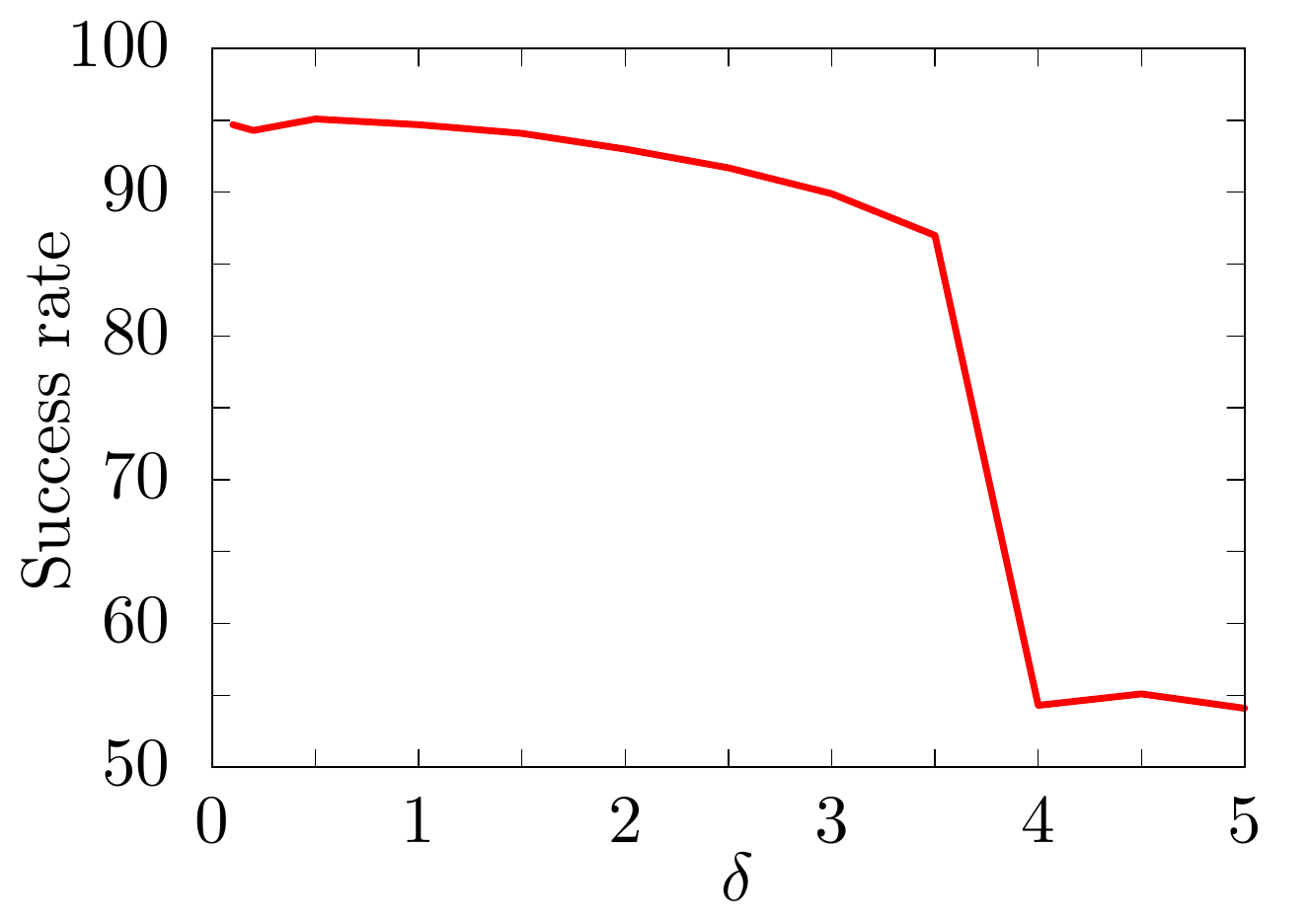}
\centering
\caption{$\delta$-dependence of success rates. Each success rate is the best one in one hundred trials.}
\label{numerical-01-05}
\end{figure}

In Fig.~\ref{numerical-01-05}, we show the $\delta$-dependence of the best success rates of the $\delta$-EM algorithm in one hundred trials.
This figure shows that the $\delta$-EM algorithm is robust for small $\delta$, but the performance decreases rapidly for large values of $\delta$.
We need to set $\delta$ small, since the critical value depends on data sets.

\subsection{Example II}

We again start with the data set used in this subsection.
The data points are also generated by the mixture of two Gaussian functions, but the parameters are different.
We set $[\pi^1, \pi^2] = [0.7, 0.3]$, $\mu^1 = [0.0, -0.5]^\intercal$, $\mu^2 = [0.0, 0.0]^\intercal$, and
\begin{align}
  \Sigma^1 &=
  \begin{bmatrix}
    1.0 & 0.0 \\
    0.0 & 1.0
  \end{bmatrix}, \\
  \Sigma^2 &=
  \begin{bmatrix}
    10.0 & 0.0 \\
    0.0 & 0.10
  \end{bmatrix}.
\end{align}
Furthermore, we draw 1000 data points from the mixture of two Gaussian functions.

\begin{figure}[t]
\begin{subfigure}[t]{0.22\textwidth}
\centering
\includegraphics[scale=0.30]{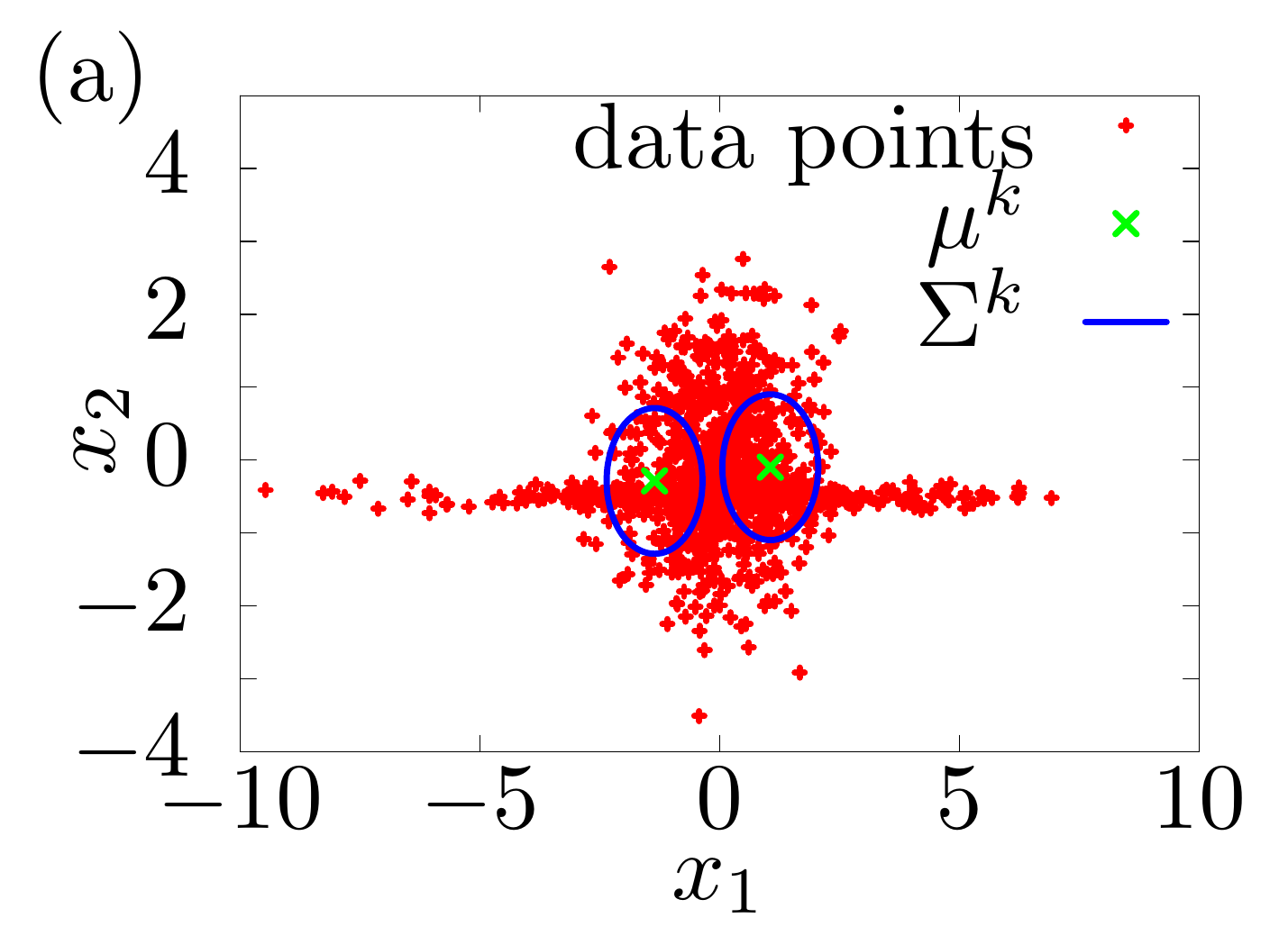}
\end{subfigure}
\begin{subfigure}[t]{0.22\textwidth}
\centering
\includegraphics[scale=0.30]{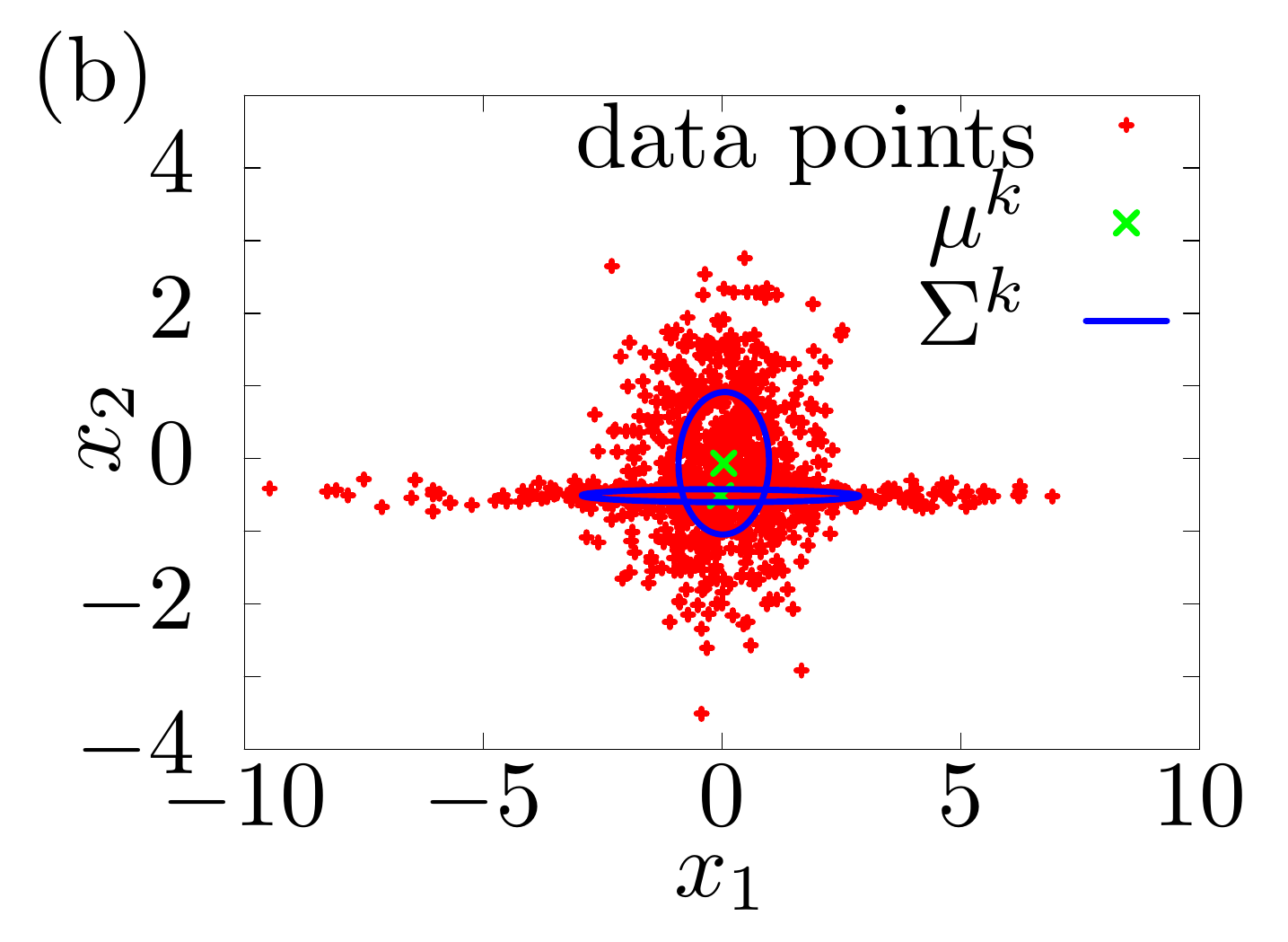}
\end{subfigure}
\caption{Pictures of estimated functions by (a)~the $\delta$-$k$-means algorithm and (b)~the $\delta$-EM algorithm.}
\label{numerical-02-04}
\end{figure}

We first show the parameters estimated by the $\delta$-$k$-means algorithm and the $\delta$-EM algorithm in Fig.~\ref{numerical-02-04}.
We have chosen the best estimates of the $\delta$-$k$-means algorithm and the $\delta$-EM algorithm in one hundred trials.
These figures represent that the $\delta$-EM algorithm outperforms the $\delta$-$k$-means algorithm.
In particular, in the case of the $\delta$-$k$-means algorithm, the covariances are fixed at the identity matrix; then, each cluster tries to exclude each other.

In Table~\ref{success-rate-02-01}, we summarize the success rates of the EM algorithm, the $\delta$-EM algorithm, the $k$-means algorithm, and the $\delta$-$k$-means algorithm.
Here, the success rate means that the ratio of the number of the successfully predicted labels to the number of the total data points.
This table shows that the $\delta$-EM algorithm works better than the $\delta$-$k$-means algorithm.
\begin{table}
\centering
\begin{tabular}{c c c c}
  \hline
  \hline
  EM & $\delta$-EM  & $k$-means & $\delta$-$k$-means \\
  \hline
  88.8 \% & 89.2 \%  & 57.9 \% & 55.4 \% \\
  \hline
  \hline
\end{tabular}
\caption{Success rates of the EM algorithm, the $\delta$-EM algorithm, the $k$-means algorithm, and the $\delta$-$k$-means algorithm. These scores are best ones in one hundred trials with randomized initial inputs.}
\label{success-rate-02-01}
\end{table}

\section{Discussions} \label{disc-01}

We here discuss the relationship between the EM algorithm with the GMM and the $k$-means algorithm.
The EM algorithm with the GMM is an extension of the $k$-means algorithm; thus we explain the two conditions that the EM algorithm with the GMM becomes identical to the $k$-means algorithm.

The first condition is that $r_t^{i, k}$ takes $1$ for a certain $k$ and $0$ otherwise.
This implies that the $k$-means algorithm is an algorithm for hard clustering, while the EM algorithm is one for soft clustering.
The second condition is that $\pi^k = 1 / K$ and $\Sigma^k = I_d$ where $I_d$ is the $d$-dimensional identity matrix for $k = 1, 2, \dots, K$.
This is the reason why the $k$-means algorithm does not explicitly deal with weights and covariance matrices.
From the viewpoint of a probability distribution, the $k$-means algorithm is an algorithm to estimate $\{ \mu_t^k \}_{k=1}^K$ in
\begin{align}
		p (x; \{ \mu^k \}_{k=1}^K) &= \frac{1}{K} \frac{1}{(2 \pi)^\frac{d}{2}} \sum_{k=1}^K e^{- \frac{1}{2} \| x - \mu^k \|^2}.
\end{align}

As shown in Sec.~\ref{sec-numerical-simulation}, the weights and the covariances of the GMM play an important role; thus, we also insist that the $\delta$-EM algorithm is a meaningful extension of the $\delta$-$k$-means algorithm.

\section{Conclusion} \label{conc-01}

In this paper, we have proposed a quantum algorithm for the EM algorithm and showed that it realize a quantum speedup compared to the classical EM algorithm.
The key idea is to generalize the distance that is minimized in the $k$-means algorithm by considering also weights and covariances.
Though we have focused on the GMM, we can generalize this condition to other mixture models.
In machine learning, the EM algorithm with the GMM is more often used than the $k$-means algorithm; thus, this work is an important step toward quantum machine learning.
The algorithm requires a QRAM oracle which has so far not been implemented in experiments yet. As a future direction we will investigate the applicability of novel superposition designs as proposed in Refs. \cite{sieberer2018programmable, dlaska2019designing}.

\section*{Acknowledgements} \label{ack-00}

H.M. thanks Clemens Dlaska for fruitful discussions.
H.M. is supported by JSPS KAKENHI Grant No. JP18J12175.
The work is supported by the Austrian Science Fund (FWF) through a START grant under Project No. Y1067-N27 and the SFB BeyondC Project No.~F7108-N38, the Hauser-Raspe foundation, and the European Union's Horizon 2020 research and innovation program under grant agreement No.~817482 PasQuanS.



\appendix

\section{Big $O$ notation} \label{sec-big-o-notation-01}

We here introduce the big $O$ notation, which is often used in computer science.
For functions $f (x)$ and $g (x)$, one writes
\begin{align}
f (x) &= O ( g (x)),
\end{align}
if and only if
\begin{align}
  \exists x_0, \exists M > 0, \ \mathrm{s.t.} \ x > x_0 \Rightarrow \| f (x) \| < M \| g (x) \|.
\end{align}
Similarly, we say
\begin{align}
f (x) &= \tilde{O} (g (x)),
\end{align}
if and only if
\begin{align}
  \exists k, f (x) &= O \Big( g (x) \ln^k (x) \Big).
\end{align}
These definitions will be utilized to describe the q-EM algorithm and to perform error analysis.

\section{Quantum preliminaries}

We provide some tools that are required for the $q$-means algorithm~\cite{Kerenidis02} in this section.
These tools are also utilized in a quantum algorithm for the EM algorithm.

\subsection{Amplitude estimation} \label{amp-est-01-01}

Here we sumarize the amplitude estimation algorithm that was proposed in Ref.~\cite{Brassard01}. Assume that we have $U_A$ given by
\begin{align}
  U_A: \Ket{0} \mapsto \sqrt{p} \Ket{\mathrm{tar}_{i, k}, 1} + \sqrt{1 - p} \Ket{\mathrm{gar}_{i, k}, 0}.
\end{align}
Then, there exists an amplitude estimation algorithm that outputs $\tilde{p}$ such that
\begin{align}
  \| \tilde{p} - p \| &\le 2 \pi \frac{\sqrt{p (1 - p)}}{P_\mathrm{ae}} + \bigg( \frac{\pi}{P_\mathrm{ae}} \bigg)^2,
\end{align}
with probability at least $8 / \pi^2$.
The algorithm perform $U_A$ $P_\mathrm{ae}$ times.
Note that, if $p = 0$, $\tilde{p} = 0$, and if $p = 0$ and $P_\mathrm{ae}$ is even, then $\tilde{p} = 0$.

Furthermore, to raise the probability to obtain a good estimate on distances, we utilize a tool in Ref.~\cite{Wiebe01}.
We make multiple copies of the amplitude estimates, apply the quantum mode evaluation algorithm proposed in Lemma~8 of Ref.~\cite{Wiebe01} in Sec.~\ref{med-eval-01-01}, and reverse the circuit to remove the garbage state. We note, that very recently an amplitude estimation algorithm without phase estimation was introduced \cite{suzuki2019amplitude}.

\subsection{Median evaluation} \label{med-eval-01-01}

The time complexity of the mode evaluation algorithm is given in Lemma~8 of Ref.~\cite{Wiebe01}. Let us summarize the main idea of this Lemma. Let $U$ be a unitary operation given by
\begin{align}
  U: \Ket{0^{\otimes n}} \mapsto \sqrt{a} \Ket{x, 1} + \sqrt{1 - a} \Ket{\mathrm{gar}, 0},
\end{align}
for $1/2 < a \le 1$ in time $T$.
Then there exists a quantum algorithm that produces a state $\Ket{\Phi}$ such that
\begin{align}
  \| \Ket{\Phi_{i, k}} - \Ket{0}^{\otimes n L} \Ket{x} \|_2 &\le \sqrt{2 \Delta},
\end{align}
for $\Delta > 0$, $1/2 < a_0 < a$ and integer $L$ in time $2T \Big\lceil \frac{\ln (1 / \Delta)}{2 (|a_0| - 1/2)^2} \Big\rceil$.

\subsection{Quantum random access memory}

In the $q$-means and $q$-EM algorithm, it is crucial to prepare data as a quantum state efficiently. To this end, we exploit the quantum random access memory (QRAM) introduced in Refs. ~\cite{Giovannetti01, Giovannetti02}. Here, the authors consider a device that performs the operation
\begin{align}
  \sum_j \psi_j \Ket{j}_\mathrm{a} \overset{\mathrm{QRAM}}{\to} \sum_j \psi_j \Ket{j}_\mathrm{a} \Ket{D_j}_\mathrm{d}. \label{QRAM-01-01}
\end{align}

We follow the application of the QRAM as in Ref.~\cite{Kerenidis02}.
Let $V_1 \in \mathbb{R}^{N \times d}$; then, there is a data structure to store the rows of $V_1$ such that the time to insert, update, or delete a single entry $v_{i, j}$ is $O (\ln^2 N)$ and a quantum algorithm on the data structure can be performed in time $O (\ln^2 N)$ that realizes the following unitaries:
\begin{align}
  \Ket{i} \Ket{0} &\mapsto \Ket{i} \Ket{v_i} \ \text{for $i \in [N]$}, \\
  \Ket{0} &\mapsto \sum_{i \in [N]} \| v_i \| \Ket{i}.
\end{align}

\subsection{Quantum linear algebra} \label{sec-qla-01}

Some useful subroutines that are used in $q$-means and $q$-EM are given as follows:
\begin{theorem} \label{theorem-qla-01-01}
Let $M \in \mathbb{R}^{d \times d}$ that satisfies $\| M \|_2 = 1$ and $x \in \mathbb{R}^d$.
If $M$ is stored in QRAM and the time to prepare $\Ket{x}$ is $T_x$, then there exist quantum algorithms that return
\begin{itemize}
  \item a state $\Ket{z}$ such that $\| \Ket{z} - \Ket{M x} \| \le \epsilon$ in time $\tilde{O} \Big( (\kappa (M) \mu (M) + T_x \kappa (M)) \ln (\epsilon^{-1}) \Big)$,
  \item a state $\Ket{z}$ such that $\| \Ket{z} - \Ket{M^{-1} x} \| \le \epsilon$ in time $\tilde{O} \Big( (\kappa (M) \mu (M) + T_x \kappa (M)) \ln (\epsilon^{-1}) \Big)$,
  \item the norm $z \in (1 + \delta) \| M x \|$ with relative error $\delta$ in time $\tilde{O} \Big( T_x \kappa (M) \mu (M) \delta^{-1} \ln (\epsilon^{-1}) \Big)$.
\end{itemize}
where
\begin{align}
  \mu (M) &\coloneqq \min_{p \in [0, 1]} \Big(\| M \|_\mathrm{F}, \sqrt{s_{2p} (M) s_{1 - 2p} (M^\intercal)} \Big), \label{def-mu-01-01} \\
  s_p (M) &\coloneqq \max_{i \in [n]} \sum_{j \in [d]} M_{i, j}^p,
\end{align}
and $\kappa (M)$ is the condition number of $M$.
\end{theorem}

\section{Inequality for error analysis} \label{sec-supplementary-calculation}

The following holds for general $\vec{a}$ and $\vec{b}$:
\begin{align}
  & \big\| \| \vec{a} \| \cdot | \vec{a} \rangle - \| \vec{b} \| \cdot | \vec{b} \rangle \big\| \nonumber \\
  & \quad \le \big\| \| \vec{a} \| \cdot | \vec{a} \rangle - \| \vec{a} \| \cdot | \vec{b} \rangle \big\| + \big\| \| \vec{a} \| \cdot | \vec{b} \rangle - \| \vec{b} \| \cdot | \vec{b} \rangle \big\| \\
  & \quad = \| \vec{a} \| \cdot \big\| | \vec{a} \rangle - | \vec{b} \rangle \big\| + (\| \vec{a} \|| - \| \vec{b} \|) \big\| | \vec{b} \rangle \big\|.
\end{align}
We used the above equation to derive Eq.~\eqref{error-bound-01-31}.

\bibliographystyle{apsrev4-1}
\bibliography{paper-quantum-algo-GMM-99-01-bib}

\begin{thebibliography}{31}%
\makeatletter
\providecommand \@ifxundefined [1]{%
 \@ifx{#1\undefined}
}%
\providecommand \@ifnum [1]{%
 \ifnum #1\expandafter \@firstoftwo
 \else \expandafter \@secondoftwo
 \fi
}%
\providecommand \@ifx [1]{%
 \ifx #1\expandafter \@firstoftwo
 \else \expandafter \@secondoftwo
 \fi
}%
\providecommand \natexlab [1]{#1}%
\providecommand \enquote  [1]{``#1''}%
\providecommand \bibnamefont  [1]{#1}%
\providecommand \bibfnamefont [1]{#1}%
\providecommand \citenamefont [1]{#1}%
\providecommand \href@noop [0]{\@secondoftwo}%
\providecommand \href [0]{\begingroup \@sanitize@url \@href}%
\providecommand \@href[1]{\@@startlink{#1}\@@href}%
\providecommand \@@href[1]{\endgroup#1\@@endlink}%
\providecommand \@sanitize@url [0]{\catcode `\\12\catcode `\$12\catcode
  `\&12\catcode `\#12\catcode `\^12\catcode `\_12\catcode `\%12\relax}%
\providecommand \@@startlink[1]{}%
\providecommand \@@endlink[0]{}%
\providecommand \url  [0]{\begingroup\@sanitize@url \@url }%
\providecommand \@url [1]{\endgroup\@href {#1}{\urlprefix }}%
\providecommand \urlprefix  [0]{URL }%
\providecommand \Eprint [0]{\href }%
\providecommand \doibase [0]{http://dx.doi.org/}%
\providecommand \selectlanguage [0]{\@gobble}%
\providecommand \bibinfo  [0]{\@secondoftwo}%
\providecommand \bibfield  [0]{\@secondoftwo}%
\providecommand \translation [1]{[#1]}%
\providecommand \BibitemOpen [0]{}%
\providecommand \bibitemStop [0]{}%
\providecommand \bibitemNoStop [0]{.\EOS\space}%
\providecommand \EOS [0]{\spacefactor3000\relax}%
\providecommand \BibitemShut  [1]{\csname bibitem#1\endcsname}%
\let\auto@bib@innerbib\@empty
\bibitem [{\citenamefont {Shor}(1999)}]{Shor01}%
  \BibitemOpen
  \bibfield  {author} {\bibinfo {author} {\bibfnamefont {P.~W.}\ \bibnamefont
  {Shor}},\ }\href@noop {} {\bibfield  {journal} {\bibinfo  {journal} {SIAM
  review}\ }\textbf {\bibinfo {volume} {41}},\ \bibinfo {pages} {303} (\bibinfo
  {year} {1999})}\BibitemShut {NoStop}%
\bibitem [{\citenamefont {Nielsen}\ and\ \citenamefont
  {Chuang}(2002)}]{Nielsen02}%
  \BibitemOpen
  \bibfield  {author} {\bibinfo {author} {\bibfnamefont {M.~A.}\ \bibnamefont
  {Nielsen}}\ and\ \bibinfo {author} {\bibfnamefont {I.}~\bibnamefont
  {Chuang}},\ }\href@noop {} {\enquote {\bibinfo {title} {Quantum computation
  and quantum information},}\ } (\bibinfo {year} {2002})\BibitemShut {NoStop}%
\bibitem [{\citenamefont {Biamonte}\ \emph {et~al.}(2017)\citenamefont
  {Biamonte}, \citenamefont {Wittek}, \citenamefont {Pancotti}, \citenamefont
  {Rebentrost}, \citenamefont {Wiebe},\ and\ \citenamefont
  {Lloyd}}]{biamonte2017quantum}%
  \BibitemOpen
  \bibfield  {author} {\bibinfo {author} {\bibfnamefont {J.}~\bibnamefont
  {Biamonte}}, \bibinfo {author} {\bibfnamefont {P.}~\bibnamefont {Wittek}},
  \bibinfo {author} {\bibfnamefont {N.}~\bibnamefont {Pancotti}}, \bibinfo
  {author} {\bibfnamefont {P.}~\bibnamefont {Rebentrost}}, \bibinfo {author}
  {\bibfnamefont {N.}~\bibnamefont {Wiebe}}, \ and\ \bibinfo {author}
  {\bibfnamefont {S.}~\bibnamefont {Lloyd}},\ }\href@noop {} {\bibfield
  {journal} {\bibinfo  {journal} {Nature}\ }\textbf {\bibinfo {volume} {549}},\
  \bibinfo {pages} {195} (\bibinfo {year} {2017})}\BibitemShut {NoStop}%
\bibitem [{\citenamefont {Lloyd}\ \emph {et~al.}(2013)\citenamefont {Lloyd},
  \citenamefont {Mohseni},\ and\ \citenamefont {Rebentrost}}]{Lloyd01}%
  \BibitemOpen
  \bibfield  {author} {\bibinfo {author} {\bibfnamefont {S.}~\bibnamefont
  {Lloyd}}, \bibinfo {author} {\bibfnamefont {M.}~\bibnamefont {Mohseni}}, \
  and\ \bibinfo {author} {\bibfnamefont {P.}~\bibnamefont {Rebentrost}},\
  }\href@noop {} {\bibfield  {journal} {\bibinfo  {journal} {arXiv preprint
  arXiv:1307.0411}\ } (\bibinfo {year} {2013})}\BibitemShut {NoStop}%
\bibitem [{\citenamefont {Schuld}\ \emph {et~al.}(2015)\citenamefont {Schuld},
  \citenamefont {Sinayskiy},\ and\ \citenamefont
  {Petruccione}}]{schuld2015introduction}%
  \BibitemOpen
  \bibfield  {author} {\bibinfo {author} {\bibfnamefont {M.}~\bibnamefont
  {Schuld}}, \bibinfo {author} {\bibfnamefont {I.}~\bibnamefont {Sinayskiy}}, \
  and\ \bibinfo {author} {\bibfnamefont {F.}~\bibnamefont {Petruccione}},\
  }\href@noop {} {\bibfield  {journal} {\bibinfo  {journal} {Contemporary
  Physics}\ }\textbf {\bibinfo {volume} {56}},\ \bibinfo {pages} {172}
  (\bibinfo {year} {2015})}\BibitemShut {NoStop}%
\bibitem [{\citenamefont {Rebentrost}\ \emph {et~al.}(2014)\citenamefont
  {Rebentrost}, \citenamefont {Mohseni},\ and\ \citenamefont
  {Lloyd}}]{Robentrost01}%
  \BibitemOpen
  \bibfield  {author} {\bibinfo {author} {\bibfnamefont {P.}~\bibnamefont
  {Rebentrost}}, \bibinfo {author} {\bibfnamefont {M.}~\bibnamefont {Mohseni}},
  \ and\ \bibinfo {author} {\bibfnamefont {S.}~\bibnamefont {Lloyd}},\ }\href
  {\doibase 10.1103/PhysRevLett.113.130503} {\bibfield  {journal} {\bibinfo
  {journal} {Phys. Rev. Lett.}\ }\textbf {\bibinfo {volume} {113}},\ \bibinfo
  {pages} {130503} (\bibinfo {year} {2014})}\BibitemShut {NoStop}%
\bibitem [{\citenamefont {Wiebe}\ \emph {et~al.}(2014)\citenamefont {Wiebe},
  \citenamefont {Kapoor},\ and\ \citenamefont {Svore}}]{Wiebe01}%
  \BibitemOpen
  \bibfield  {author} {\bibinfo {author} {\bibfnamefont {N.}~\bibnamefont
  {Wiebe}}, \bibinfo {author} {\bibfnamefont {A.}~\bibnamefont {Kapoor}}, \
  and\ \bibinfo {author} {\bibfnamefont {K.}~\bibnamefont {Svore}},\
  }\href@noop {} {\bibfield  {journal} {\bibinfo  {journal} {arXiv preprint
  arXiv:1401.2142}\ } (\bibinfo {year} {2014})}\BibitemShut {NoStop}%
\bibitem [{\citenamefont {Dunjko}\ \emph {et~al.}(2016)\citenamefont {Dunjko},
  \citenamefont {Taylor},\ and\ \citenamefont {Briegel}}]{dunjko2016quantum}%
  \BibitemOpen
  \bibfield  {author} {\bibinfo {author} {\bibfnamefont {V.}~\bibnamefont
  {Dunjko}}, \bibinfo {author} {\bibfnamefont {J.~M.}\ \bibnamefont {Taylor}},
  \ and\ \bibinfo {author} {\bibfnamefont {H.~J.}\ \bibnamefont {Briegel}},\
  }\href@noop {} {\bibfield  {journal} {\bibinfo  {journal} {Physical review
  letters}\ }\textbf {\bibinfo {volume} {117}},\ \bibinfo {pages} {130501}
  (\bibinfo {year} {2016})}\BibitemShut {NoStop}%
\bibitem [{\citenamefont {Bishop}(2007)}]{Bishop01}%
  \BibitemOpen
  \bibfield  {author} {\bibinfo {author} {\bibfnamefont {C.}~\bibnamefont
  {Bishop}},\ }\href@noop {} {\enquote {\bibinfo {title} {Pattern recognition
  and machine learning (information science and statistics), 1st edn. 2006.
  corr. 2nd printing edn},}\ } (\bibinfo {year} {2007})\BibitemShut {NoStop}%
\bibitem [{\citenamefont {Murphy}(2012)}]{Murphy01}%
  \BibitemOpen
  \bibfield  {author} {\bibinfo {author} {\bibfnamefont {K.~P.}\ \bibnamefont
  {Murphy}},\ }\href@noop {} {\emph {\bibinfo {title} {Machine learning: a
  probabilistic perspective}}}\ (\bibinfo  {publisher} {MIT press},\ \bibinfo
  {year} {2012})\BibitemShut {NoStop}%
\bibitem [{\citenamefont {Kerenidis}\ \emph {et~al.}(2018)\citenamefont
  {Kerenidis}, \citenamefont {Landman}, \citenamefont {Luongo},\ and\
  \citenamefont {Prakash}}]{Kerenidis02}%
  \BibitemOpen
  \bibfield  {author} {\bibinfo {author} {\bibfnamefont {I.}~\bibnamefont
  {Kerenidis}}, \bibinfo {author} {\bibfnamefont {J.}~\bibnamefont {Landman}},
  \bibinfo {author} {\bibfnamefont {A.}~\bibnamefont {Luongo}}, \ and\ \bibinfo
  {author} {\bibfnamefont {A.}~\bibnamefont {Prakash}},\ }\href@noop {}
  {\bibfield  {journal} {\bibinfo  {journal} {arXiv preprint arXiv:1812.03584}\
  } (\bibinfo {year} {2018})}\BibitemShut {NoStop}%
\bibitem [{\citenamefont {Dempster}\ \emph {et~al.}(1977)\citenamefont
  {Dempster}, \citenamefont {Laird},\ and\ \citenamefont {Rubin}}]{Dempster01}%
  \BibitemOpen
  \bibfield  {author} {\bibinfo {author} {\bibfnamefont {A.~P.}\ \bibnamefont
  {Dempster}}, \bibinfo {author} {\bibfnamefont {N.~M.}\ \bibnamefont {Laird}},
  \ and\ \bibinfo {author} {\bibfnamefont {D.~B.}\ \bibnamefont {Rubin}},\
  }\href@noop {} {\bibfield  {journal} {\bibinfo  {journal} {Journal of the
  Royal Statistical Society, Series B}\ }\textbf {\bibinfo {volume} {39}},\
  \bibinfo {pages} {1} (\bibinfo {year} {1977})}\BibitemShut {NoStop}%
\bibitem [{\citenamefont {Miyahara}\ and\ \citenamefont
  {Tsumura}(2016)}]{Miyahara03}%
  \BibitemOpen
  \bibfield  {author} {\bibinfo {author} {\bibfnamefont {H.}~\bibnamefont
  {Miyahara}}\ and\ \bibinfo {author} {\bibfnamefont {K.}~\bibnamefont
  {Tsumura}},\ }in\ \href@noop {} {\emph {\bibinfo {booktitle} {American
  Control Conference (ACC), 2016}}}\ (\bibinfo {year} {2016})\BibitemShut
  {NoStop}%
\bibitem [{\citenamefont {Miyahara}\ \emph {et~al.}(2016)\citenamefont
  {Miyahara}, \citenamefont {Tsumura},\ and\ \citenamefont
  {Sughiyama}}]{Miyahara04}%
  \BibitemOpen
  \bibfield  {author} {\bibinfo {author} {\bibfnamefont {H.}~\bibnamefont
  {Miyahara}}, \bibinfo {author} {\bibfnamefont {K.}~\bibnamefont {Tsumura}}, \
  and\ \bibinfo {author} {\bibfnamefont {Y.}~\bibnamefont {Sughiyama}},\ }in\
  \href@noop {} {\emph {\bibinfo {booktitle} {Decision and Control (CDC), 2016
  IEEE 55th Conference on}}}\ (\bibinfo {organization} {IEEE},\ \bibinfo {year}
  {2016})\ pp.\ \bibinfo {pages} {4674--4679}\BibitemShut {NoStop}%
\bibitem [{\citenamefont {Miyahara}\ \emph {et~al.}(2017)\citenamefont
  {Miyahara}, \citenamefont {Tsumura},\ and\ \citenamefont
  {Sughiyama}}]{Miyahara05}%
  \BibitemOpen
  \bibfield  {author} {\bibinfo {author} {\bibfnamefont {H.}~\bibnamefont
  {Miyahara}}, \bibinfo {author} {\bibfnamefont {K.}~\bibnamefont {Tsumura}}, \
  and\ \bibinfo {author} {\bibfnamefont {Y.}~\bibnamefont {Sughiyama}},\ }\href
  {http://stacks.iop.org/1742-5468/2017/i=11/a=113404} {\bibfield  {journal}
  {\bibinfo  {journal} {Journal of Statistical Mechanics: Theory and
  Experiment}\ }\textbf {\bibinfo {volume} {2017}},\ \bibinfo {pages} {113404}
  (\bibinfo {year} {2017})}\BibitemShut {NoStop}%
\bibitem [{\citenamefont {Miyahara}\ and\ \citenamefont
  {Sughiyama}(2018)}]{Miyahara06}%
  \BibitemOpen
  \bibfield  {author} {\bibinfo {author} {\bibfnamefont {H.}~\bibnamefont
  {Miyahara}}\ and\ \bibinfo {author} {\bibfnamefont {Y.}~\bibnamefont
  {Sughiyama}},\ }\href@noop {} {\bibfield  {journal} {\bibinfo  {journal}
  {Physical Review A}\ }\textbf {\bibinfo {volume} {98}},\ \bibinfo {pages}
  {022330} (\bibinfo {year} {2018})}\BibitemShut {NoStop}%
\bibitem [{\citenamefont {Kerenidis}\ \emph {et~al.}(2019)\citenamefont
  {Kerenidis}, \citenamefont {Luong},\ and\ \citenamefont
  {Prakash}}]{Kerenidis05}%
  \BibitemOpen
  \bibfield  {author} {\bibinfo {author} {\bibfnamefont {I.}~\bibnamefont
  {Kerenidis}}, \bibinfo {author} {\bibfnamefont {A.}~\bibnamefont {Luong}}, \
  and\ \bibinfo {author} {\bibfnamefont {A.}~\bibnamefont {Prakash}},\
  }\href@noop {} {\bibfield  {journal} {\bibinfo  {journal} {(Private
  communication)}\ } (\bibinfo {year} {2019})}\BibitemShut {NoStop}%
\bibitem [{\citenamefont {Brassard}\ \emph {et~al.}(2002)\citenamefont
  {Brassard}, \citenamefont {Hoyer}, \citenamefont {Mosca},\ and\ \citenamefont
  {Tapp}}]{Brassard01}%
  \BibitemOpen
  \bibfield  {author} {\bibinfo {author} {\bibfnamefont {G.}~\bibnamefont
  {Brassard}}, \bibinfo {author} {\bibfnamefont {P.}~\bibnamefont {Hoyer}},
  \bibinfo {author} {\bibfnamefont {M.}~\bibnamefont {Mosca}}, \ and\ \bibinfo
  {author} {\bibfnamefont {A.}~\bibnamefont {Tapp}},\ }\href@noop {} {\bibfield
   {journal} {\bibinfo  {journal} {Contemporary Mathematics}\ }\textbf
  {\bibinfo {volume} {305}},\ \bibinfo {pages} {53} (\bibinfo {year}
  {2002})}\BibitemShut {NoStop}%
\bibitem [{Note1()}]{Note1}%
  \BibitemOpen
  \bibinfo {note} {This algorithm has no specific name in Ref.~\cite {Wiebe01},
  and it is called median evaluation in Ref.~\cite {Kerenidis02}. But this
  algorithm realize majority voting; so we call it mode
  evaluation.}\BibitemShut {Stop}%
\bibitem [{\citenamefont {Chakraborty}\ \emph {et~al.}(2018)\citenamefont
  {Chakraborty}, \citenamefont {Gily{\'e}n},\ and\ \citenamefont
  {Jeffery}}]{Chakraborty01}%
  \BibitemOpen
  \bibfield  {author} {\bibinfo {author} {\bibfnamefont {S.}~\bibnamefont
  {Chakraborty}}, \bibinfo {author} {\bibfnamefont {A.}~\bibnamefont
  {Gily{\'e}n}}, \ and\ \bibinfo {author} {\bibfnamefont {S.}~\bibnamefont
  {Jeffery}},\ }\href@noop {} {\bibfield  {journal} {\bibinfo  {journal} {arXiv
  preprint arXiv:1804.01973}\ } (\bibinfo {year} {2018})}\BibitemShut {NoStop}%
\bibitem [{\citenamefont {Kerenidis}\ and\ \citenamefont
  {Luongo}(2018)}]{Kerenidis03}%
  \BibitemOpen
  \bibfield  {author} {\bibinfo {author} {\bibfnamefont {I.}~\bibnamefont
  {Kerenidis}}\ and\ \bibinfo {author} {\bibfnamefont {A.}~\bibnamefont
  {Luongo}},\ }\href@noop {} {\bibfield  {journal} {\bibinfo  {journal} {arXiv
  preprint arXiv:1805.08837}\ } (\bibinfo {year} {2018})}\BibitemShut {NoStop}%
\bibitem [{\citenamefont {Hoeffding}(1994)}]{Hoeffding01}%
  \BibitemOpen
  \bibfield  {author} {\bibinfo {author} {\bibfnamefont {W.}~\bibnamefont
  {Hoeffding}},\ }in\ \href@noop {} {\emph {\bibinfo {booktitle} {The Collected
  Works of Wassily Hoeffding}}}\ (\bibinfo  {publisher} {Springer},\ \bibinfo
  {year} {1994})\ pp.\ \bibinfo {pages} {409--426}\BibitemShut {NoStop}%
\bibitem [{\citenamefont {Cornell}\ and\ \citenamefont
  {Sastry}(2015)}]{Cornell01}%
  \BibitemOpen
  \bibfield  {author} {\bibinfo {author} {\bibfnamefont {D.}~\bibnamefont
  {Cornell}}\ and\ \bibinfo {author} {\bibfnamefont {S.}~\bibnamefont
  {Sastry}},\ }\href@noop {} {\  (\bibinfo {year} {2015})}\BibitemShut
  {NoStop}%
\bibitem [{Note2()}]{Note2}%
  \BibitemOpen
  \bibinfo {note} {In the the $\delta $-$k$-means algorithm, we added Gaussian
  noise whose mean is 0 and variance is 0.01 to each element of all the
  centroids. In the M step of the $\delta $-EM algorithm, we added Gaussian
  noise whose mean is 0 and variance is 0.01 to $\pi ^k$ and each element of
  $\mu ^k$, and add Gaussian noise whose mean is 0 and variance is 0.001 to
  each element of $\Sigma ^k$. When $\pi $ is not normalized, we normalized
  $\pi $. When $\Sigma ^k$ is not symmetric, we symmetrize $\Sigma ^k$ by
  $(\Sigma ^k + (\Sigma ^k)^\intercal ) / 2$. When the estimated $\Sigma ^k$
  has a negative eigenvalue $\sigma ^*$, we make all eigenvalues positive by
  adding $| \sigma ^* | I^k$ where $I^k$ is the $k \times k$ identity
  matrix}\BibitemShut {NoStop}%
\bibitem [{Note3()}]{Note3}%
  \BibitemOpen
  \bibinfo {note} {In Fig.~\ref {numerical-01-04}, we draw the blue ellipses by
  moving $\theta $ $(0 \le \theta < 2 \pi )$ in $f (\theta ) \mathrel {\mathop
  :}\mathrel {\mkern -1.2mu}=\mu ^k + \protect \sqrt {\sigma _1^k} e_1^k
  \protect \qopname \relax o{cos}(\theta ) + \protect \sqrt {\sigma _2^k} e_2^k
  \protect \qopname \relax o{sin}\theta $ where $\sigma _1^k$ and $\sigma _2^k$
  are the eigenvalues of $\Sigma ^k$, and $e_1^k$ and $e_2^k$ are the
  corresponding eigenvectors of $\Sigma ^k$, respectively, for $k = 1, 2$. In
  Fig.~\ref {numerical-02-04}, we also draw blue ellipses by the same
  procedure.}\BibitemShut {Stop}%
\bibitem [{Note4()}]{Note4}%
  \BibitemOpen
  \bibinfo {note} {The hidden variable in the GMM means the index of the
  Gaussian functions. Note that the GMM has the symmetry on swapping the
  indices; then, we use the maximum value on swapping.}\BibitemShut {Stop}%
\bibitem [{\citenamefont {Sieberer}\ and\ \citenamefont
  {Lechner}(2018)}]{sieberer2018programmable}%
  \BibitemOpen
  \bibfield  {author} {\bibinfo {author} {\bibfnamefont {L.~M.}\ \bibnamefont
  {Sieberer}}\ and\ \bibinfo {author} {\bibfnamefont {W.}~\bibnamefont
  {Lechner}},\ }\href@noop {} {\bibfield  {journal} {\bibinfo  {journal}
  {Physical Review A}\ }\textbf {\bibinfo {volume} {97}},\ \bibinfo {pages}
  {052329} (\bibinfo {year} {2018})}\BibitemShut {NoStop}%
\bibitem [{\citenamefont {Dlaska}\ \emph {et~al.}(2019)\citenamefont {Dlaska},
  \citenamefont {Sieberer},\ and\ \citenamefont
  {Lechner}}]{dlaska2019designing}%
  \BibitemOpen
  \bibfield  {author} {\bibinfo {author} {\bibfnamefont {C.}~\bibnamefont
  {Dlaska}}, \bibinfo {author} {\bibfnamefont {L.~M.}\ \bibnamefont
  {Sieberer}}, \ and\ \bibinfo {author} {\bibfnamefont {W.}~\bibnamefont
  {Lechner}},\ }\href@noop {} {\bibfield  {journal} {\bibinfo  {journal}
  {Physical Review A}\ }\textbf {\bibinfo {volume} {99}},\ \bibinfo {pages}
  {032342} (\bibinfo {year} {2019})}\BibitemShut {NoStop}%
\bibitem [{\citenamefont {Suzuki}\ \emph {et~al.}(2019)\citenamefont {Suzuki},
  \citenamefont {Uno}, \citenamefont {Raymond}, \citenamefont {Tanaka},
  \citenamefont {Onodera},\ and\ \citenamefont
  {Yamamoto}}]{suzuki2019amplitude}%
  \BibitemOpen
  \bibfield  {author} {\bibinfo {author} {\bibfnamefont {Y.}~\bibnamefont
  {Suzuki}}, \bibinfo {author} {\bibfnamefont {S.}~\bibnamefont {Uno}},
  \bibinfo {author} {\bibfnamefont {R.}~\bibnamefont {Raymond}}, \bibinfo
  {author} {\bibfnamefont {T.}~\bibnamefont {Tanaka}}, \bibinfo {author}
  {\bibfnamefont {T.}~\bibnamefont {Onodera}}, \ and\ \bibinfo {author}
  {\bibfnamefont {N.}~\bibnamefont {Yamamoto}},\ }\href@noop {} {\bibfield
  {journal} {\bibinfo  {journal} {arXiv preprint arXiv:1904.10246}\ } (\bibinfo
  {year} {2019})}\BibitemShut {NoStop}%
\bibitem [{\citenamefont {Giovannetti}\ \emph
  {et~al.}(2008{\natexlab{a}})\citenamefont {Giovannetti}, \citenamefont
  {Lloyd},\ and\ \citenamefont {Maccone}}]{Giovannetti01}%
  \BibitemOpen
  \bibfield  {author} {\bibinfo {author} {\bibfnamefont {V.}~\bibnamefont
  {Giovannetti}}, \bibinfo {author} {\bibfnamefont {S.}~\bibnamefont {Lloyd}},
  \ and\ \bibinfo {author} {\bibfnamefont {L.}~\bibnamefont {Maccone}},\
  }\href@noop {} {\bibfield  {journal} {\bibinfo  {journal} {Physical review
  letters}\ }\textbf {\bibinfo {volume} {100}},\ \bibinfo {pages} {160501}
  (\bibinfo {year} {2008}{\natexlab{a}})}\BibitemShut {NoStop}%
\bibitem [{\citenamefont {Giovannetti}\ \emph
  {et~al.}(2008{\natexlab{b}})\citenamefont {Giovannetti}, \citenamefont
  {Lloyd},\ and\ \citenamefont {Maccone}}]{Giovannetti02}%
  \BibitemOpen
  \bibfield  {author} {\bibinfo {author} {\bibfnamefont {V.}~\bibnamefont
  {Giovannetti}}, \bibinfo {author} {\bibfnamefont {S.}~\bibnamefont {Lloyd}},
  \ and\ \bibinfo {author} {\bibfnamefont {L.}~\bibnamefont {Maccone}},\
  }\href@noop {} {\bibfield  {journal} {\bibinfo  {journal} {Physical Review
  A}\ }\textbf {\bibinfo {volume} {78}},\ \bibinfo {pages} {052310} (\bibinfo
  {year} {2008}{\natexlab{b}})}\BibitemShut {NoStop}%
\end{thebibliography}%

\end{document}